\documentclass[10pt,journal,compsoc]{IEEEtran}
\ifCLASSOPTIONcompsoc
  \usepackage[nocompress]{cite}
\else
  \usepackage{cite}
\fi
\ifCLASSINFOpdf
   \usepackage[pdftex]{graphicx}
\else
\fi
\usepackage{bm}
\usepackage{amssymb}
\usepackage{amsfonts}
\usepackage[margin=2.5cm]{geometry}
\usepackage{graphicx}
\usepackage{multirow}
\usepackage{url}
\usepackage[table]{xcolor}
\usepackage{amsmath}
\usepackage{algorithm}
\usepackage{algorithmic}

\graphicspath{{./}{images/}}
\usepackage{array}  
\newcolumntype{H}{>{\setbox0=\hbox\bgroup}c<{\egroup}@{}} 

\begin{document}

\title{One-dimensional discrete-time quantum walks with general coin}
\author{Mahesh N. Jayakody, Chandrakala Meena and  Priodyuti Pradhan 
\IEEEcompsocitemizethanks{\IEEEcompsocthanksitem Mahesh N. Jayakody is associated with Faculty of Engineering, Bar-Ilan University, Ramat Gan, Israel. E-mail: jadmn.jayakody@gmail.com
\IEEEcompsocthanksitem Chandrakala Meena is associated with the Council of Scientific and Industrial Research -- National Chemical Laboratory (CSIR-NCL) Pune, Dr Homi Bhabha Rd, Ward No. 8, NCL Colony, Pashan, Pune, Maharashtra 411008. E-mail: meenachandrakala@gmail.com
\IEEEcompsocthanksitem Priodyuti Pradhan is associated with the Complex Network Dynamics Lab, Mathematics Department, Bar-Ilan University, Ramat Gan, Israel. E-mail: priodyutipradhan@gmail.com
}

}

\markboth{Journal of \LaTeX\ Class Files,~Vol.~14, No.~8, \today}%
{Shell \MakeLowercase{\textit{et al.}}: Bare Demo of IEEEtran.cls for Computer Society Journals}

\IEEEtitleabstractindextext{%
\begin{abstract}
Quantum walk (QW) is the quantum analog of the random walk. QW is an integral part of the development of numerous quantum algorithms. Hence, an in-depth understanding of QW helps us to grasp the quantum algorithms. We revisit the one-dimensional discrete-time QW and discuss basic steps in detail by incorporating the most general coin operator. 
We investigate the impact of each parameter of the general coin operator on the probability distribution of the quantum walker. We show that by tuning the parameters of the general coin, one can regulate the probability distribution of the walker. We provide an algorithm for the one-dimensional quantum walk driven by the general coin operator. The study conducted on general coin operator also includes the popular coins -- Hadamard, Grover, and Fourier coins.
\end{abstract}

\begin{IEEEkeywords}
Quantum walk, Hadamard coin, Grover coin, Fourier coin, Quantum entanglement. 
\end{IEEEkeywords}}

\maketitle

\IEEEdisplaynontitleabstractindextext
\IEEEpeerreviewmaketitle
\IEEEraisesectionheading{\section{Introduction}\label{sec:introduction}}

\IEEEPARstart{W}e have entered the quantum technology era that puts quantum mechanical principles into practice \cite{Mohseni1}. Quantum mechanics is a fundamental theory in physics that describes the physical properties of nature at the scale of atoms and subatomic particles. Making use of the properties in the realm of atoms and subatomic particles, quantum technology provides a promising enhancement in our day-to-day utilities \cite{Dowling1}. For instance, the atomic clock used in the satellites is a direct application of quantum mechanics, which synchronizes the time with high precision over the globe to provide a reliable GPS navigation system to smartphone users \cite{Bresson}. In recent times, quantum technologies have started empowering fast computing, secure communication, accurate imaging in healthcare, and advancing radar technology in military services \cite{Acin}. In this article, we explore a frequently used tool in quantum technologies, quantum walk (QW) \cite{jayakody_2019, Chandrashekar_qw_2020}, which has diverse applications in the development of algorithms \cite{Wang}. 

QWs are quantum analogs of classical random walks \cite{random_walk_quantum_walk_2020, Quantum_algorithms_2016}. The very first conceptual proposal, which could be recognized as a QW, was published in a seminal work by R. P. Feynman related to quantum mechanical computers \cite{Feynman}. Nonetheless, it is generally accepted that the first paper which explicitly defines the QWs was published by Aharonov et al. in 1993 \cite{Aharonov}. Principally, QWs contribute to theoretical and practical improvements in quantum algorithms \cite{Kempe}, and in quantum computing \cite{Childs_2013}. For instance, QW improves cryptoanalysis \cite{qw_cryptography_2020}, complex network analysis \cite{qw_complex_networks_2019}, secure data transfer in 5G internet of things (IoT) communications and in wireless networking with edge computing platforms \cite{qw_IoT_2020,qw_5G_communication_2020}. Algorithms based on QWs provide exponential speedup for some oracle problems compared to any classical algorithms \cite{Childs_2007}. Moreover, QW based algorithms give polynomial speedups over classical algorithms for problems such as element distinctness \cite{Ambainis}, and in triangle finding \cite{Magniez}. QWs are used to model transport in biological systems \cite{Hoyer}, physical phenomena such as Anderson localization \cite{Xue} and also in understanding the topological phases \cite{Kitagawa_2012}. QWs have also been utilized to develop a single particle graph isomorphism algorithm that can successfully distinguish all pairs of graphs, including all strongly regular graphs with up to $64$ vertices \cite{Douglas}. 

There are two broad classes of QWs known as discrete-time QWs (DTQW) and continuous-time QWs (CTQW), each of which has significant distinctions in their mathematical formalism \cite{Wang}. One of their significant differences is that DTQW is defined on a discrete-time domain, and in contrast, CTQW is defined on a continuous-time domain. However, it has been shown that DTQWs can be transformed into a CTQWs under certain conditions \cite{Strauch_2006}. Here, we focus on the DTQWs on the integer line. Such QWs are referred to as one-dimensional discrete-time QWs or simply as QWs on a line. It has the simplest structure of a QW and helps us to understand the higher dimensional walks on graphs \cite{qw_graphs_2002}. One-dimensional QWs have three major components -- a) coin operator, b) shift operator, and c) initial coin state. Starting from an initial state and by successively applying the coin and the shift operators, we get the probability distribution of a quantum walker. It is a well-known fact that the final behavior of the walker for the Hadamard coin depends on the initial coin state \cite{Venegas-Andraca}. In the Hadamard walk, we get a symmetric probability distribution for an unbiased initial coin state, and for a biased initial coin state, the final probability distribution becomes asymmetric. Even though studies of QWs on a line have been done extensively, the coin operators used in those studies are mostly limited to the Hadamard, Grover, and Fourier coins, and an exploration of the general coin operator is still lacking from the engineering perspective \cite{Venegas-Andraca}. Our contribution can be summarized as follows
\begin{itemize}
    \item We explore the probability distribution of the one-dimensional discrete-time QWs with the general coin.
    
    \item We show that by tuning the general coin's parameters, we can make the final probability distribution asymmetric even for an unbiased initial coin state.
    
    \item Our analysis reveals that the rotation and a single phase parameter of the general coin play an essential role in controlling the probability distribution of the walker.
    
    \item We show that our numerical results are in good agreement with that of the analytical derivations.
    
    \item We provide an algorithm for the one-dimensional QWs on line.
    
    \item We give some intuitions of the quantum entanglement in the context of QWs.
    
    \item We present steps of the QW as simple as possible and illustrate the matrix notations. 
\end{itemize}
We fabricate the article as follows, in section \ref{mathematical_framework}, we begin our discussion with the classical random walk and subsequently discuss the QW from the quantum mechanical point of view in order to understand how the concepts of the classical random walk are implemented in the quantum regime. In section \ref{matrix_notation_algorithm}, we transform the abstract quantum mechanical notations into concrete matrix form for the implementation. Further, in section 4, we discuss the results \& discussion and provide an algorithm for the one-dimensional QW driven by the most general coin. Finally, in section 5, we conclude our results and discuss some open problems for further research.
\begin{table*}[t]
\begin{center}
\begin{tabular}{|l|p{8cm}|} 
\hline
Symbols & Description \\
\hline
$\mathbb{Z}$ & integer numbers\\
\hline
$\mathbb{C}$ & complex numbers\\
\hline
$\otimes$ & tensor product\\ 
\hline
$|\cdot \rangle $ &  column vector or ket\\ 
\hline
$\langle \cdot |\cdot \rangle$ & inner product  \\ 
\hline
$\mathcal{H}_{\bm{c}}$ & Hilbert space associated to coin space\\ 
\hline
$\mathcal{H}_{\bm{x}}$ & Hilbert space associated to position space\\ 
\hline
$\mathcal{H}=\mathcal{H}_{\bm{x}} \otimes \mathcal{H}_{\bm{c}}$ & Hilbert space associated to walker state space \\ 
\hline
$\mathcal{C}$, ${\bf C}$ & coin operator and associated matrix
\\
\hline
$\mathcal{S}$, ${\bf S}$ & shift operator and associated matrix
\\
\hline
${|\bm{x} \rangle}_{x=-\infty}^{\infty}$ & Position states \\ 
\hline
$\alpha_{x}(t) |{\bf H} \rangle$ +$\beta_{x}(t) |{\bf T} \rangle$ & Coin states as superposition of two states $|{\bf H}\rangle$ and $|{\bf T}\rangle$\\
\hline
$|\psi(x,t) \rangle$ & state of the particle at position $\bm{x}$ at time $t$. \\
\hline
$|\psi_t \rangle=\sum_{\bm{x}=-\infty}^{\infty}|\bm{\psi}(\bm{x},t)\rangle$ & state of the particle at time $t$. \\
\hline
$P(\bm{x},t)$& probability of finding the particle at time t with position $\bm{x}$.\\
\hline
$P(t)=\sum_{\bm{x}=-\infty}^{\infty} P(\bm{x},t)$ & total probability\\
\hline
$|\alpha|^2$, $|\beta|^2$ & probability at $t=0$ \\
\hline
\end{tabular}
\end{center}
\caption{List of symbols and notations.}
\end{table*}

\section{Mathematical Framework of QWs}\label{mathematical_framework}
Let us consider the one-dimensional discrete-time classical random walk (RW) in which a particle on the integer line jumps either to the right or left depending upon the outcome of a coin toss \cite{random_walk_quantum_walk_2020}. At a given time, the motion of the particle on the integer line consists of two sequential processes, the first being the tossing of the coin and the second being the shifting of the particle that depends upon the outcome of the coin. In other words, RW is a random process that captures the dynamical behavior of the particle. For the sake of simplicity, we assume the particle is placed at the origin. We can then move the particle on the integer line for a desired number of steps by using the coin toss and the shifting rule. Suppose we record the position of the particle at each step. Then by using the recorded data, we can generate the probability distribution of the walker and which converges to the famous Bell curve \cite{Weiss_1983}. 

\subsection{Modeling quantum walks using the rules of quantum mechanics}
It is possible to define an analog of RW in the quantum regime as well. For that, we need to define the quantum version of the RW by incorporating the rules of quantum mechanics. We can identify two components of an RW -- (a) the integer line on which the particle is moving and (b) the outcome of the coin toss. Our task is to define the quantum version of the RW by transforming these components into the quantum regime. 

According to the postulates of quantum mechanics, we can analyze the dynamical behavior of a physical system by associating a complex Hilbert space to the system \cite{Harris_1995}. And we represent any state of the physical system by an element of the complex Hilbert space.
For instance, let us consider an isolated atom comprised of a single proton and an electron having four distinct energy states. This can be considered as a physical system. 
Now, we attach a complex Hilbert space to our system. We represent each energy state by using the basis elements of the Hilbert space. Since we have four distinct energy states, we need a basis set with four elements. Thus, we associate a Hilbert space spanned by this basis set to our system to study the dynamics in terms of quantum mechanics. Note that, for four-level energy states, we need four-dimensional Hilbert space. However, an atom can have an infinite number of energy states, and hence to represent such a system, we need infinite-dimensional Hilbert space.  
Further, consider another isolated physical system that consists of a single particle having two possible spin states. For instance, an isolated electron can be considered a  physical system with two possible spin states called 'spin up' and 'spin down'. Like in the previous case, to study the dynamics of this single particle system, we need to attach a Hilbert space that spans from a basis set with two elements. Next, consider a scenario in which one needs to study the behavior of both the systems, the isolated atom, and the isolated single particle, together as a whole system. Then one needs to combine the Hilbert spaces attached to each system in a reasonable manner. The operation of tensor products ($\otimes$) of Hilbert spaces comes into the picture to resolve this problem.

In connection to the example of the isolated atom and the isolated single particle, let us discuss how to develop the mathematical model for QWs in detail. As mentioned earlier, at a given time, a particle moving in an RW has a specific position on the integer line and a specific coin outcome. To define the quantum version of it, we attach the position and the coin outcome of the quantum particle with two suitable Hilbert spaces.
Like in the example of an atom and an electron, we can think  
positions of the particle as the energy states of the atom and spin states of the electron as the outcome of the coin toss. Since the set of integers is infinite, the quantum walker can occupy an infinite number of position states. Therefore, the Hilbert space attached to the position must have an infinite dimension. That is, the basis set of the Hilbert space attached to the position must be infinite. Let $\mathcal{H}_{\bm{x}}$ be the Hilbert space spanned by the basis set $\{|\bm{x}\rangle \}_{x\in\mathbb{Z}}$. Further, the coin state has only two possible outcomes (head and tail).  Hence, we can attach the coin state to a Hilbert space that spans from a basis set containing two elements. Let $\mathcal{H}_{\bm{c}}$ be the Hilbert space spanned by the basis set $\{|{\bf H}\rangle, |{\bf T}\rangle \}$. 
Note that the coin has two states. A quantum system that has two states is termed as a two-level system \cite{Thaller}. In the context of quantum information, any two-level system is considered as a qubit \cite{Thaller}. Hence, one can view the coin of the quantum walker as a qubit.

By attaching the position state and the coin state of the quantum particle  
to $\mathcal{H}_{\bm{x}}$ and $\mathcal{H}_{\bm{c}}$, one can represent the wave function or state ($|{\bf \psi}(\bm{x},t) \rangle \in \mathcal{H}=\mathcal{H}_{\bm{x}} \otimes \mathcal{H}_{\bm{c}}$) of the quantum particle at any position $\bm{x}$ and time $t$ as 
\begin{equation}\label{Eq1}
    |\bm{\psi}(\bm{x},t)\rangle = (\alpha_{\bm{x}}(t)|{\bf H}\rangle+\beta_{\bm{x}}(t)|{\bf T}\rangle)\otimes |\bm{x}\rangle
\end{equation}
where the positions state is represented by $|\bm{x}\rangle$ and the coin state is represented by the superposition of two basis states of $|{\bf H}\rangle$ and $|{\bf T}\rangle$ as \cite{Wang} 
\begin{equation}\label{coin_state}
\alpha_{\bm{x}}(t)|{\bf H}\rangle+\beta_{\bm{x}}(t)|{\bf T}\rangle    
\end{equation}
where $\alpha_{\bm{x}}(t)$ and $\beta_{\bm{x}}(t)$ are some complex numbers. Then the probability of finding the quantum walker at position $\bm{x}$ at time $t$ can be calculated as 
\begin{equation}\label{Eq8}
    P(\bm{x},t)=\langle \bm{\psi}(\bm{x},t)|\bm{\psi}(\bm{x},t)\rangle = |\alpha_{\bm{x}}(t)|^2+|\beta_{\bm{x}}(t)|^2
\end{equation} 
under the condition $P(\bm{x},t) \leq 1$. Further, we can find the total wave function or state of the QW at time $t$ from Eq. (\ref{Eq1}) as 
\begin{equation}\label{Eq9}
\begin{split}
    |\bm{\psi}_t\rangle&=\sum_{\bm{x}=-\infty}^{\infty}|\bm{\psi}(\bm{x},t)\rangle\\
    &=\sum_{\bm{x}=-\infty}^{\infty}(\alpha_{\bm{x}}(t)|{\bf H}\rangle+\beta_{\bm{x}}(t)|{\bf T}\rangle)\otimes |\bm{x} \rangle
\end{split}    
\end{equation} 
Now, with the help of the total wave function, we can calculate the total probability distribution of the quantum walker as follows
\begin{equation}\label{total_probability}
     P_t=\langle \bm{\psi}_t|\bm{\psi}_t\rangle = \sum_{\bm{x}=-\infty}^{\infty}{|\alpha_{\bm{x}}(t)|^2+|\beta_{\bm{x}}(t)|^2}=1
\end{equation}
Our interest lies in understanding the total probability distribution in detail, which captures the dynamics of the quantum walker. From Eq. (\ref{Eq9}), we can see that particle can occupy infinite number of places. However, for real-world implementation, we transform the infinite number of position states into a finite number. 

\subsection{Quantum mechanical operators}
In quantum mechanics, the time evolution of any physical process in a closed system is defined in terms of a unitary transformation that alters a given initial state ($|\bm{\psi}_i\rangle$) of the system to a final state ($|\bm{\psi}_{f}\rangle$) in a reversible manner \cite{Bowers}. Thus, one needs to utilize unitary operators in describing how the state of a quantum mechanical system changes with time. Therefore, when implementing a QW it is essential to define a unitary operators that mimic the sequential processes of coin toss and shifting. The  unitary operator that perform the coin toss operation is referred as the general coin operator for an one-dimensional QW and it is defined as 
\begin{equation}\label{coin_operator}
\begin{split}
\mathcal{C} &= \cos \theta|{\bf H}\rangle \langle {\bf H}|+e^{i\phi_1}\sin \theta|{\bf H}\rangle \langle {\bf T}|+\\
      &e^{i\phi_2}\sin \theta|{\bf T}\rangle \langle {\bf H}|
      -e^{i(\phi_1+\phi_2)}\cos \theta|{\bf T}\rangle \langle {\bf T}|
\end{split}      
\end{equation}
where $\theta\in [0,2\pi)$ and $\phi_1,\phi_2\in [0,\pi)$ \cite{general_coin_operator_2003}. We refer $\theta$ as {\em rotation} and $\phi_1$ \& $\phi_2$ as  {\em phase} parameters of general coin operator. 
For instance, when $\theta=45^{\circ}$, $\phi_1=\phi_2=0^{\circ}$, the coin becomes the famous Hadamard coin operator. Further, for $\theta=90^{\circ}$, $\phi_1=\phi_2=0^{\circ}$, we obtain the Grover coin and when $\theta=45^{\circ}$, $\phi_1=\phi_2=90^{\circ}$, we get the Fourier coin \cite{Gratsea}. Most of the previous research have focused only on these three coins and for the other values of $\theta$, $\phi_1$, and $\phi_2$ the walker probability distribution remains elusive. In this article, we explicitly explore the general coin operator for all possible combinations of rotation and phase parameters and investigate the total probability distribution. Now, let us define the unitary operator that perform the shift operation as
\begin{equation}\label{shift_operator}
\begin{split}
   \mathcal{S}&= |{\bf H}\rangle \langle {\bf H}|\otimes \biggl(\sum_{x=-\infty}^{\infty}{|\bm{x+1}\rangle\langle \bm{x}|}\biggr)+\\
   &|{\bf T}\rangle \langle {\bf T}| \otimes \biggl(\sum_{x=-\infty}^{\infty}{|\bm{x-1}\rangle\langle \bm{x}|}\biggr)
\end{split}   
\end{equation}
The mathematical operation $\otimes$ is termed as tensor product. In plain language, the meaning of tensor products is that the operations on each space are executed separately. Note that the operator $\mathcal{S}$ treats the coin states and the position state separately.
Therefore, in comparison with RW, the unitary operator $\mathcal{C}$ is used for the quantum coin toss, and the unitary operator $\mathcal{S}$ is used to shift the particle on the integer line. Hence, a single-step progression of the QW on a line can be written as a sequential process in which the quantum coin is tossed at first, and then the walker is moved conditionally upon the outcome of the coin.
\begin{figure*}[!ht]
\centerline{\includegraphics[width=4.8in, height=2.8in]{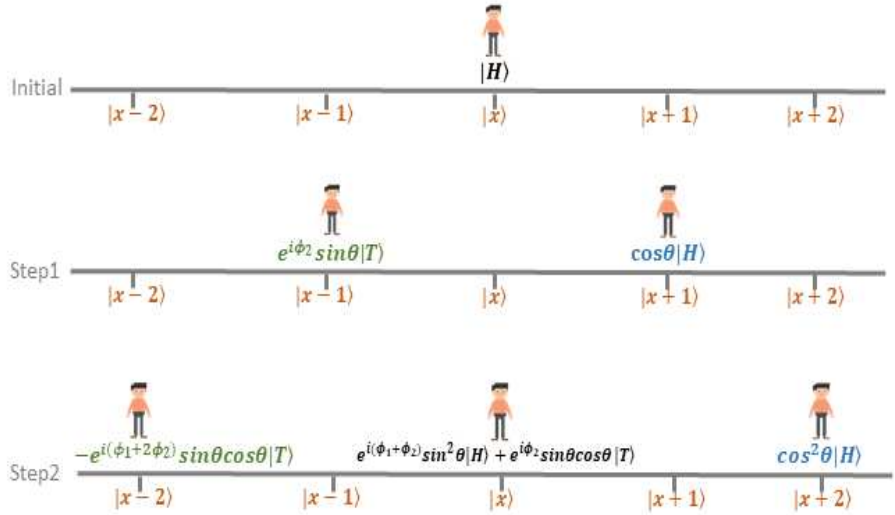}}
\caption {Progression of QW for the first and second steps initiating from the state $|{\bf H}\rangle \otimes |\bm{x}\rangle$. Here,  $|\bm{x-2}\rangle,\ldots|\bm{x+2}\rangle$ are the position states of the walker on the integer line and $|{\bf H}\rangle$ is the initial coin state. We observe that by applying the coin and shift operator on the initial coin state $|{\bf H}\rangle$ at position state $|\bm{x}\rangle$, walker can reside two places simultaneously ($|\bm{x-1}\rangle$ and $|\bm{x+1}\rangle$) with probabilities $\sin^2\theta$ and $\cos^2\theta$ respectively. After the second step, walker resides on $|\bm{x-2}\rangle$, $|\bm{x}\rangle$ and $|\bm{x+2}\rangle$ with probabilities 
$\sin^2\theta\cos^2\theta$,  
$\sin^4\theta+\sin^2\theta\cos^2\theta$ and $\cos^2\theta$ respectively.}
\label{fig2}
\end{figure*}
The unitary operator that corresponds to a single-step evolution of the QW on a line is given by
\begin{equation}\label{unitary_operator}
 {\mathcal{U}}=\mathcal{S}(\mathcal{C} \otimes \mathcal{I}) 
\end{equation}
where $\mathcal{I}$ is identity operator. Let us see an example of how the evolution of the QW occurs under the operation of $\mathcal{U}$. Suppose at time $t=0$ the position of the particle is $\bm{x}$, and the initial state of the coin is $\alpha_{\bm{x}}(0)|{\bf H}\rangle+\beta_{\bm{x}}(0)|{\bf T}\rangle$ (Eq. (\ref{coin_state})). Then the state of the particle is represented as a tensor product of coin and position states and we denote it as
\begin{equation}\label{zero_state}
|\bm{\psi}_0 \rangle=|\bm{\psi}(\bm{x},0)\rangle = \biggl(\alpha_{\bm{x}}(0)|{\bf H}\rangle+\beta_{\bm{x}}(0)|{\bf T}\rangle\biggr) \otimes |\bm{x}\rangle 
\end{equation}
where $|\alpha_{\bm{x}}(0)|^2+|\beta_{\bm{x}}(0)|^2=1$. If we consider 
$|\alpha_{\bm{x}}(0)|^2=|\beta_{\bm{x}}(0)|^2$ in the above expression,  we get an unbiased initial coin state. In contrast, when $|\alpha_{\bm{x}}(0)|^2 \neq |\beta_{\bm{x}}(0)|^2$, we get a biased initial coin state. Our main results consider only the unbiased initial coin state. However, for the sake of simplicity, in this section we use the biased initial coin state where $|\beta_{\bm{x}}(0)|^2=0$. Hence, the initial state of the particle in Eq. (\ref{zero_state}) can be rewritten as \begin{equation}\label{initial_state}
|\bm{\psi_0}\rangle=|{\bf H}\rangle \otimes |\bm{x}\rangle
\end{equation}
which says that the walker is residing at position $\bm{x}$ with the coin state of $Head$. Now, we apply $\mathcal{U}$ on the initial state in Eq. (\ref{initial_state}) and we get (detail in SI Eq. S1)
\begin{equation}\label{Eq5}
\begin{split}
|\bm{\psi}_{1}\rangle &=\mathcal{U}|\bm{\psi}_{0}\rangle\\ 
&=\cos \theta|{\bf H}\rangle  \otimes |\bm{x+1}\rangle + e^{i\phi_2}\sin \theta|{\bf T}\rangle  \otimes |\bm{x-1}\rangle
\end{split}
\end{equation}
Initially the quantum walker was at $|\bm{x}\rangle$ position. However, after the first step quantum walker occupies two positions $|\bm{x-1}\rangle$ and $|\bm{x+1}\rangle$ where $|\cos \theta|^2$, and $|e^{i\phi_2}\sin \theta|^2$ are the probabilities of finding the walker at those positions respectively. Further, we apply $\mathcal{U}$ on $|\bm{\psi}_{1}\rangle$ and get $|\bm{\psi}_{2}\rangle$ as (detail in SI Eq. S2)
\begin{equation}\label{Eq6}
\begin{split}
|\bm{\psi}_{2}\rangle &= \mathcal{U}|\bm{\psi}_{1}\rangle\\ 
&=\cos^2 \theta|{\bf H}\rangle  \otimes |\bm{x+2}\rangle+\\
&\biggl(e^{i(\phi_1+\phi_2)}\sin^2 \theta|{\bf H}\rangle+e^{i\phi_2}\sin \theta \cos \theta|{\bf T}\rangle\biggr) \otimes  |\bm{x}\rangle\\
&-e^{i(\phi_1+2\phi_2)}\sin \theta \cos \theta|{\bf T}\rangle\otimes |\bm{x}-2\rangle
\end{split}
\end{equation}
 One can observe that after the second step quantum walker occupies three position states $|\bm{x}-2\rangle$, $|\bm{x}\rangle$, and $|\bm{x}+2\rangle$ where $|\cos^2 \theta|^2$, $|e^{i(\phi_1+\phi_2)}\sin^2 \theta|^2+|e^{i\phi_2}\sin \theta \cos \theta|^2$ and $|-e^{i(\phi_1+2\phi_2)}\sin \theta \cos \theta|^2$ are the probabilities of finding the walker at those positions respectively.
Fig. 1 illustrates the progression of QW for the first two steps. In general, we represent the state of the walker at time $t$ as
\begin{equation}\label{Unitary_operator}
 |\bm{\psi}_{t}\rangle={\mathcal{U}}^{t}|\bm{\psi}_0\rangle
\end{equation}
Note that in QWs, we always need to consider the state of the coin before it is tossed. However, in the classical random walks, we never pay attention to this matter. In classical random walks, we never worry about checking whether we keep $Head$ or $Tail$ on the top side before we toss the coin. What we do is we keep any side of the coin ($Head$ or $Tail$) on the top side and toss the coin. Then, depending on the outcome, we shift the particle. However, in QWs, a coin toss is done by applying the coin operator on the current coin state of the particle. Hence, {\em different initial coin states give different outcomes}. This is one of the subtle differences between QWs and RWs.

\section{Matrix representations}\label{matrix_notation_algorithm}
Any two Hilbert spaces whose orthonormal bases have the same cardinality are isomorphic \cite{Hunter}. This allows us to map the position Hilbert space ($\mathcal{H}_{\bm{x}}$) and the coin Hilbert space ($\mathcal{H}_{\bm{c}}$) into the corresponding matrix Hilbert space.

\subsection {QW in finite dimensional Hilbert space}
\begin{figure}[!ht]
\centerline{\includegraphics[width=80mm, height=25mm]{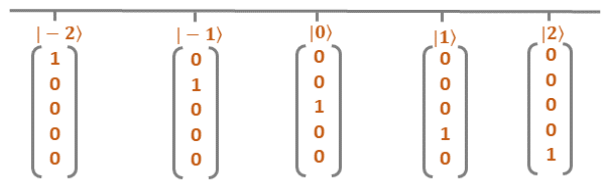}}
\caption {One-to-one mapping of the position states ($\{|\bm{-2}\rangle,\ |\bm{-1}\rangle,\ |\bm{0}\rangle,\ |\bm{1}\rangle, |\bm{2}\rangle\}$) to the standard basis elements of five dimensional matrix Hilbert space. The first basis vector is attached to the left most corner position and accordingly other basis vectors are attached to the position states.}
\label{fig3}
\end{figure}

\begin{figure}[htb]
\centerline{\includegraphics[width=3.2in, height=1.5in]{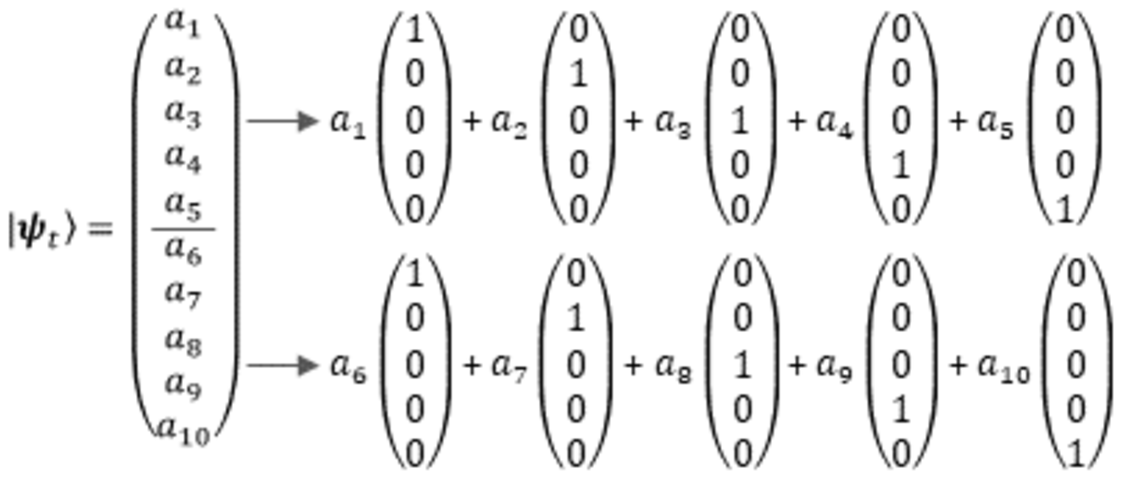}}
\caption {Decomposing a state ($|\bm{\psi}_t\rangle$) of the QW. First five elements of $|\bm{\psi}_t\rangle$ contains the coefficients corresponding to the coin state ($|{\bf H}\rangle$) and the next five elements contain the coefficients corresponding to the coin state ($|{\bf T}\rangle$). By rewriting first block and second block separately as a linear combination of basis set given in Fig. (\ref{fig3}) one can identify the location where the coin coefficients reside.}
\label{column_vector_decomposition}
\end{figure}

As discussed earlier, the position Hilbert space attached to the QW has an infinite dimension. Hence, when mapping the $\mathcal{H}_{\bm{x}}$ into the matrix Hilbert space, we need infinite-dimensional Hilbert space. However, employing an infinite-dimensional Hilbert space give rise to practical issues in the implementations. We resolve this problem by fixing the number of position states to a finite value. Let $N$ be an arbitrary positive integer. Hence, we represent the set of position states $\{|\bm{-N}\rangle,\hdots,|\bm{0}\rangle,\hdots,|\bm{N}\rangle\}$ on the integer line where the walker can reside. This can be done by mapping a set of position states to the standard basis set of $n=(2N+1)$ dimensional Hilbert space. When assigning the basis elements to position states, we follow a certain order. As a rule of thumb, we set the state $|\bm{-N}\rangle$ to the basis element of $(1\;\; 0\;\; \cdots\;\; 0)^{T}$ and then map rest of the position states accordingly. For example, if $N=2$, we have $5$ position states. We represent these position states using the standard basis of $5$-dimensional Hilbert space (Fig. \ref{fig3}). Similarly, we map the coin Hilbert space ($\mathcal{H}_{\bm{c}}$) into its corresponding matrix Hilbert space. Since there are only two coin states, we need a 2-dimensional Hilbert space to represent $\mathcal{H}_{\bm{c}}$. We map coin states $|{\bf H}\rangle$ and $|{\bf T}\rangle$ to the standard bases $(1\;\; 0)^{T}$ and $(0\;\; 1)^{T}$
respectively. Now, we express the state of the walker in Eqs. (\ref{initial_state}-\ref{Eq6}) 
in column representation (SI section 2) as 
follows where $|\bm{x}\rangle=|\bm{0}\rangle$
and we get 
\begin{equation}\label{Eq11}
    \begin{gathered}
    |\bm{\psi}_{0}\rangle
    =\begin{pmatrix}
  0 \\
  0 \\
  1 \\
  0 \\
  0 \\
  0 \\
  0 \\
  0 \\
  0 \\
  0 \\
\end{pmatrix},\;\;
|\bm{\psi}_{1}\rangle
=\begin{pmatrix}
  0 \\
  0 \\
  0 \\
  \cos \theta \\
  0 \\
  0 \\
  e^{i\phi_2}\sin \theta \\
  0 \\
  0 \\
  0 \\
\end{pmatrix}
\end{gathered}
\end{equation}
\begin{equation}\label{Eq13}
\begin{gathered}
|\bm{\psi}_{2}\rangle =\begin{pmatrix}
  0 \\
  0 \\
  e^{i(\phi_1+\phi_2)}\sin^2 \theta \\
  0 \\
  \cos^2 \theta \\
  -e^{i(\phi_1+2\phi_2)}\sin \theta \cos \theta  \\
  0 \\
  e^{i\phi_2}\sin \theta \cos \theta\\
  0 \\
  0\\
\end{pmatrix}
\end{gathered}
\end{equation}
By looking at the element of column vector representation of the state vector in Eqs. (\ref{Eq11}-\ref{Eq13}),
we can find the position of the quantum particle and also the probability of finding the particle at a certain position. We can identify that the first five elements of each column vectors (first block) hold the coefficients correspond to $|{\bf H}\rangle$. The next five elements of each column vector (second block) correspond to $|{\bf T}\rangle$. Furthermore, by representing the linear combination form of the basis for the two blocks of each column vector separately, we can get the information of particle's position and probability of being at that position (Fig. \ref{column_vector_decomposition}). Now we can also easily compare the elements of the column vectors and the coefficients of the graphical representation given in (Fig. \ref{fig2}).

Moreover, from the QW state vector, we can also observe an interesting phenomenon, i.e., quantum entanglement, which is a unique type of correlation shared between coin and position states \cite{entanglement}. It is important to note that the shift operator (Eq. \ref{shift_operator}) creates the correlation between the coin and position states. Consider the three QW states given in Eqs. (\ref{Eq11}-\ref{Eq13}). Although the walk is initialized in such a way that the coin and position states are separable ($|\bm{\psi_0}\rangle=|{\bf H}\rangle \otimes |\bm{x}\rangle$), after the first and second steps, the coin and position states of the walk are no longer separable but entangled (details in SI). Hence, QWs become a potential testing platform to understand quantum entanglement. 

\subsection {Operators of QW in Matrix Hilbert space}
In the above, we discuss about the column representation of the state vector in terms of basis elements of Hilbert spaces. Here, we use the chosen bases to explicitly construct the matrix corresponding to coin and shift operators and denote as {\bf C} and {\bf S}. We consider $|{\bf H}\rangle=(1\;\;0)^{T}$ and $|{\bf T}\rangle=(0\;\;1)^{T}$ in Eq. (\ref{coin_operator}), and get
\begin{equation}\label{coin_matrix}
{\bf C}=
\left( {\begin{array}{cc}
\cos \theta & e^{i\phi_1}\sin \theta \\
e^{i\phi_2}\sin \theta & -e^{i(\phi_1+\phi_2)}\cos \theta \\
\end{array}}\right)
 \end{equation}
where $\theta\in [0,2\pi)$ and $\phi_1,\phi_2\in [0,\pi)$. Hence, for different values of $\theta$, $\phi_1$ and $\phi_2$, we get different coins. Next, we consider the shift operator $\mathcal{S}$ given in (\ref{shift_operator}). It contains two components that belong to coin space and position space. The terms $|{\bf H}\rangle\langle {\bf H}|$ and $|{\bf T}\rangle\langle {\bf T}|$ are applied on the coin state. The terms $\sum_x{ |\bm{x+1}\rangle\langle \bm{x}|}$ and $\sum_x{ |\bm{x-1}\rangle\langle \bm{x}|}$ are applied on the position states. Let us write the component of $\mathcal{S}$ that applied on the position states as $\mathcal{M}=\sum_x{ |\bm{x+1}\rangle\langle \bm{x}|}$. Now, take the conjugate transpose of the operator $\mathcal{M}$ and we get $\mathcal{M}^\dagger=\sum_x{ |\bm{x}\rangle\langle \bm{x+1}|}$. By changing the variable $\bm{x}\rightarrow \bm{x-1}$ we can write $\mathcal{M}^\dagger=\sum_x{ |\bm{x-1}\rangle\langle \bm{x}|}$. Since, the position states are infinite, practically it is inconvenient to represent the operators $\mathcal{M}$ and $\mathcal{M^\dagger}$ as block matrices. Hence, we fix the number of position states to $\{|\bm{-N}\rangle,\hdots,|\bm{0}\rangle,\hdots,|\bm{N}\rangle\}$ where $N$ be a finite positive integer. Then the standard block matrix of the operator $\mathcal{M}$ can be defined as
\begin{equation}\nonumber
\begin{gathered}
    {\bf M}= \begin{pmatrix}
  0 & 0 & 0 & \hdots & 0 & 1 \\
  1 & 0 & 0 & \hdots & 0 & 0 \\
  0 & 1 & 0 & \hdots & 0 & 0 \\
  \vdots & \vdots & \vdots & \hdots & \vdots & \vdots   \\
  0 & 0 & 0 & \hdots & 1 & 0 \\
\end{pmatrix}_{(2N+1)\times (2N+1)} 
\end{gathered}
\end{equation}
As all the elements of {\bf M} are real numbers, ${\bf M}^\dagger={\bf M}^{T}$.
Hence, using the matrix form of $\mathcal{M}$ and $\mathcal{M}^\dagger$, one can express shift operator in (\ref{shift_operator}) as block matrix form 
\begin{equation}\nonumber
\begin{split}
{\bf S}&=\begin{pmatrix}
  1 & 0 \\
  0 & 0 \\
\end{pmatrix}\otimes\begin{pmatrix}
  0 & 0 & 0 & \hdots & 0 & 1 \\
  1 & 0 & 0 & \hdots & 0 & 0 \\
  0 & 1 & 0 & \hdots & 0 & 0 \\
  \vdots & \vdots & \vdots & \hdots & \vdots & \vdots   \\
  0 & 0 & 0 & \hdots & 1 & 0 \\
\end{pmatrix}\\ 
&+\begin{pmatrix}
  0 & 0 \\
  0 & 1 \\
\end{pmatrix}\otimes \begin{pmatrix}
  0 & 1 & 0 & \hdots & 0 & 0 \\
  0 & 0 & 1 & \hdots & 0 & 0 \\
  0 & 0 & 0 & \hdots & 0 & 0 \\
  \vdots & \vdots & \vdots & \hdots & \vdots & \vdots   \\
  1 & 0 & 0 & \hdots & 0 & 0 
\end{pmatrix}
\end{split}
 \end{equation}
having $(4N+2) \times (4N+2)$ dimensions. From the above explicit matrix formulation, we can also express the evolution of the quantum walk for a given initial state as follows,
\begin{equation}\label{unitary_eavlution_matrix}
|\bm{\psi}_t\rangle = {\bf U}^{t}|\bm{\psi}_0\rangle    
\end{equation}
where ${\bf U}={\bf S}({\bf C} \otimes {\bf I})$ is the transition matrix (details in SI section 4). Note that position states are denoted by $\{|\bm{-N}\rangle,\hdots,|\bm{0}\rangle,\hdots,|\bm{N}\rangle\}$, coin state are denoted by $\{|{\bf H}\rangle,|{\bf T}\rangle\}$ and the state of the system in different time is demoted by $\{|\bm{\psi}_0\rangle,|\bm{\psi}_1\rangle,\ldots,|\bm{\psi}_t\rangle\}$ and $\mathcal{H}_{\bm{x}} \in \mathbb{C}^{2N+1}$, $\mathcal{H}_{\bm{c}} \in \mathbb{C}^{2}$ and $\mathcal{H} \in \mathbb{C}^{4N+2}$, respectively. 

\section{Results and Discussion}
In this section, we explore the probability distribution of the QW by employing a systematic investigation. The motion of the quantum walker on a line depends upon five parameters -- two parameters ($\alpha\ \text{and} \ \beta$) from the initial coin state (Eq. (\ref{coin_state})) and three parameters from the general coin operator ($\theta$, $\phi_1$, and $\phi_2$) (Eq. (\ref{coin_operator})). By changing these five parameters,  we can regulate the shape of the total probability distribution of the walker (Eq. (\ref{total_probability})). It is a known fact that by changing the initial coin state parameters of QW driven by the Hadamard coin, one can change the probability distribution of the walker (SI Fig. S1). However, we have very little knowledge about how the total probability distribution changes when we vary the parameters of the general coin operator while keeping a fixed initial coin state. In this study, we fix the initial coin state to an unbiased state and analyze the impact of the general coin parameters on the probability distribution of the QW for a finite number of steps. Our aim is to identify the impact of these parameters ($\theta$, $\phi_1$, and $\phi_2$) on the total probability distribution of the quantum walker, which can help to regulate the dynamics. 

\subsection{Initial coin state of quantum walker}
Here, we consider the general form of the initial state of the quantum walker in Eq. (\ref{zero_state}) as  
\begin{equation}\label{general_initial_state}
|\bm{\psi}_0\rangle= (\alpha|{\bf H}\rangle+\beta|{\bf T}\rangle) \otimes |{\bf 0}\rangle    
\end{equation}
where $\alpha \in \mathbb{C}$, $\beta \in \mathbb{C}$ are the initial coefficient value at position $|\bm{0}\rangle$ at time $t=0$. The meaning of this initial state is that, in the beginning, we place the quantum walker at $|{\bf 0}\rangle$ position state with the probabilities $|\alpha|^2$ and $|\beta|^2$ of having $|{\bf H}\rangle$ and $|{\bf T}\rangle$ states respectively. By choosing different values for $\alpha$ and $\beta$, we can define an infinite number of initial states. However, in this study we focus on the initial states of the form 
\begin{equation}\label{unbiased_inital_state}
|\bm{\psi}_0\rangle=\biggl(\frac{1}{\sqrt{2}}|{\bf H} \rangle- \frac{i}{\sqrt{2}}|{\bf T}\rangle \biggr)\otimes |{\bf 0}\rangle
\end{equation}
Here, the probability of having $|{\bf H}\rangle$ and $|{\bf T}\rangle$ in initial coin state is equal ($|\alpha|^2=1/2$ and $|\beta|^2=1/2$) i.e., an unbiased initial state. Now, we explore the total probability distribution for the general coin (Eq. (\ref{coin_operator})) with all possible combinations of $\theta$, $\phi_1$ and $\phi_2$. However, for the simplicity first we vary $\theta$ and $\phi_1=\phi_2=0^{\circ}$ remain fixed to zero.

\begin{figure*}[!ht]
\centerline{\includegraphics[width=4.6in, height=4.3in]{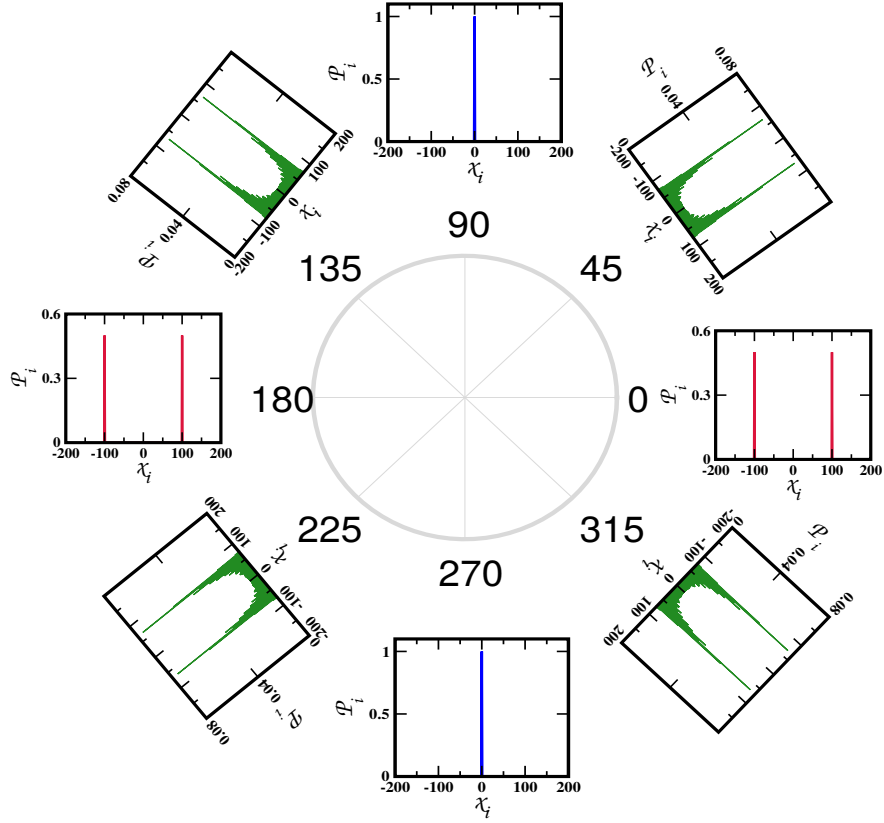}}
\caption {Probability distributions of QW for different $\theta \in [0,2\pi)$ values and under the conditions of $\phi_1=0^{\circ}$, $\phi_2=0^{\circ}$ and $t=100$.}
\label{prob_dist_1}
\end{figure*}

\subsection{Impact of rotational parameter on QW} \label{subsection:impact_of_theta}
Starting from the initial state in Eq. (\ref{unbiased_inital_state}) and we apply coin and shift operators (Eqs. (\ref{coin_operator}) and (\ref{shift_operator})). As $\theta$ varies in general coin operator from $0^{\circ}$ to $360^{\circ}$, we obtain different 
probability distributions of the walker (Fig. \ref{prob_dist_1}). 
For instance, when $\theta=90^{\circ},270^{\circ}$ the particle gets localized to the initial position. Further, for the values of $\theta=0^{\circ}\; \text{or}\; 180^{\circ}$, the distribution has two peaks with the same height, that is, the quantum walker gets localized to two extreme positions with an equal probability.
On the other hand, when $\theta=45^{\circ}$ and $\theta=225^{\circ}$, the walker can reside with non-zero probabilities in between the two extreme points. Further, for the pairs of $\theta$ and $\theta+\pi$, we get the mirror image of the same distribution. In other words, for different $\theta$ values, we get different total probability distributions of the walker. 

To understand the observations described above, we check the coin and the first few states of the walker. Recall that all the graphs (Fig. \ref{prob_dist_1}) are plotted for different $\theta$ values while keeping $\phi_1=0^{\circ}$ and $\phi_2=0^{\circ}$.
Hence, for $\theta=90^{\circ},270^{\circ}$ and $\phi_1=\phi_2=0^{\circ}$ the coins (Eq. (\ref{coin_matrix})) take the form
\begin{equation}\label{C_1_coin}
    {\bf C}_1 
      =\left( {\begin{array}{cc}
0 & 1 \\
1 & 0 \\
\end{array}}\right),\;\;
{\bf C}_2 
=-\left( {\begin{array}{cc}
0 & 1 \\
1 & 0 \\
\end{array}}\right)
\end{equation}
One can observe that the coins in (\ref{C_1_coin}) differs only by a factor of minus sign. This minus sign act as an overall phase factor. In probability calculation such overall phase factor has no contribution. Thus, we get the same effect from both the coins. Now, writing first few states of the walker using the coin in (\ref{C_1_coin}), we have
\begin{equation}\label{states_when_theta=90/270}
\begin{split} 
|\bm{\psi}_0\rangle &= \biggr(\alpha|{\bf H}\rangle+\beta|{\bf T}\rangle\biggr)  |{\bf 0}\rangle \\ 
|\bm{\psi}_1\rangle &={\bf U}|\bm{\psi}_0\rangle= \alpha|{\bf T}\rangle|\bm{-1}\rangle+\beta|{\bf H}\rangle  |{\bf 1}\rangle \\ 
|\bm{\psi}_2\rangle &={\bf U}|\bm{\psi}_1\rangle= \biggr(\alpha|{\bf H}\rangle+\beta|{\bf T}\rangle\biggr)  |{\bf 0}\rangle =|\bm{\psi}_0\rangle\\ 
|\bm{\psi}_3\rangle &={\bf U}|\bm{\psi}_0\rangle= \alpha|{\bf T}\rangle|\bm{-1}\rangle+\beta|{\bf H}\rangle  |{\bf 1}\rangle =|\bm{\psi}_1\rangle\\
&\vdots\\
|\bm{\psi}_{t}\rangle &={\bf U}|\bm{\psi}_{1}\rangle= \biggr(\alpha|{\bf H}\rangle+\beta|{\bf T}\rangle\biggr) =|\bm{\psi}_0\rangle
\end{split}
\end{equation}
From (\ref{states_when_theta=90/270}) we can observe that when $\theta=90^{\circ}\; \text{or}\;270^{\circ}$ the coefficients of the initial states $\alpha$ and $\beta$ toggles between the coin states $|{\bf H}\rangle$ and $|{\bf T}\rangle$ and as a result the walker is confined to the position states $|\bm{0}\rangle$
when $t$ is even with the probability distribution,
\begin{equation}\nonumber
P_{t} =P(\bm{0},t)= |\alpha|^2+|\beta|^2  
\end{equation}
Thus, for an unbiased initial coin state (Eq. (\ref{unbiased_inital_state})) with an even time step, we can observe localization of the walker at $|\bm{0}\rangle$ position when $\theta=90^{\circ}\;\text{or}\; 270^{\circ}$. 

Another observation we made in Fig. \ref{prob_dist_1} is that when $\theta=0^{\circ},180^{\circ}$ we get two sharp peaks with equal heights at left and right most corners of the plot. Following the previous approach, for $\theta=0^{\circ},180^{\circ}$ and $\phi_1=\phi_2=0^{\circ}$ the coins takes the following form
\begin{equation}\label{C_3_coin} 
    {\bf C}_3 =\left( {\begin{array}{cc}
1 & 0 \\
0 & -1 
\end{array}}\right),\;\;
{\bf C}_4 =-\left( {\begin{array}{cc}
1 & 0 \\
0 & -1 \\
\end{array}}\right)
\end{equation}
Like in the previous case, coins in (\ref{C_3_coin}) differs only by a factor of minus sign and we get the same effect from both the coins in probability calculation. By writing the states using the coins given in (\ref{C_3_coin}) we get
\begin{equation}\label{states_when_theta=0/180}
\begin{split} 
|\bm{\psi}_0\rangle &= \biggr(\alpha|{\bf H}\rangle+\beta|{\bf T}\rangle\biggr)  |{\bf 0}\rangle \\ 
|\bm{\psi}_1\rangle &= \alpha|{\bf H}\rangle|\bm{1}\rangle+(-1)\beta|{\bf T}\rangle  |{\bf -1}\rangle \\ 
|\bm{\psi}_2\rangle &= \alpha|{\bf H}\rangle|\bm{2}\rangle+ (-1)^2\beta|{\bf T}\rangle  |{\bf -2}\rangle \\  
|\bm{\psi}_3\rangle &= \alpha|{\bf H}\rangle|\bm{3}\rangle+ (-1)^3\beta|{\bf T}\rangle  |{\bf -3}\rangle \\ 
&\vdots\\
|\bm{\psi}_t\rangle &= \alpha|{\bf H}\rangle|\bm{N}\rangle+ (-1)^t\beta|{\bf T}\rangle  |{\bf -N}\rangle 
\end{split}
\end{equation} 
where $N=t$. Thus, the total probability of finding the walker at the left and right most corner positions, 
\begin{equation}\nonumber
P_t =P(\bm{N},t)+P(\bm{-N},t)
= |\alpha|^2+|\beta|^2    
\end{equation}
Thus, when $\theta=0^{\circ},180^{\circ}$ we get two sharp peaks with equal heights at left and right most corners. Furthermore, for coins (${\bf C}_i$ and ${\bf C}_j$) corresponding to $\theta$ and $\theta+\pi$ differ only by a factor of minus sign. That is, both coins differ only by an overall phase factor (${\bf C}_i=-{\bf C}_j$). We know from Eqs. (\ref{total_probability}) and (\ref{unitary_eavlution_matrix}) 
\begin{equation}\nonumber
\begin{split}
P_{t} &= \langle \bm{\psi}_t|\bm{\psi}_t\rangle\\
&=\langle \bm{\psi}_0|({\bf U}^t)^\dagger{\bf U}^t|\bm{\psi}_0\rangle\\
&=\langle \bm{\psi}_0|\biggl({\bf (S(-C \otimes I))}^t\biggr)^\dagger{\bf (S(-C \otimes I))}^t|\bm{\psi}_0\rangle\\
&=(-1)^{2t}\langle \bm{\psi}_0|\biggl({\bf (S(C \otimes I))}^t\biggr)^\dagger{\bf (S(C \otimes I))}^t|\bm{\psi}_0\rangle\\
&=\langle \bm{\psi}_0|\biggl({\bf (S(C \otimes I))}^t\biggr)^\dagger{\bf (S(C \otimes I))}^t|\bm{\psi}_0\rangle
\end{split}
\end{equation}
As mentioned earlier, in probability calculation, such an overall phase factor has no contribution. Thus, we get the same effect from both coins. Hence, we get the same distribution for the pairs of $\theta$ and $\theta+\pi$ (Fig. \ref{prob_dist_1}). 

\begin{figure}[!ht]
\centerline{\includegraphics[width=3.3in, height=2.1in]{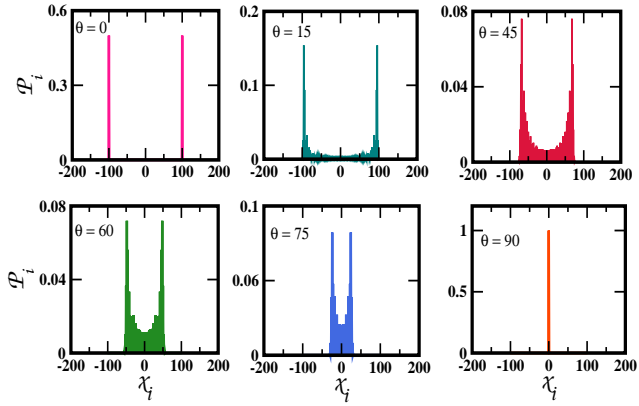}}
\caption {Probability distributions of QW when $\theta$ varies form $0^{\circ}$ to $90^{\circ}$ where $\phi_1=\phi_2=0^{\circ}$ and $t=100$. We observe that as $\theta$ value increases the walker will be localized to the initial position (${|\bm{0}\rangle}$).}
\label{plot_qw_1}
\end{figure}

Finally, we magnify the $\theta$ value in between $0^{\circ}$ and $90^{\circ}$ and observe the total probability distribution of the quantum walker (Fig. \ref{plot_qw_1}). One can observe that when $\theta=0^{\circ}$ the quantum walker tends to stay at the opposite end positions with equal probabilities. However, when the value of $\theta$ increases, the probability of finding the walker in the left and rightmost position decreases. At the same time, as the $\theta$ increases, the distribution tends to shrink towards the initial position, and when $\theta=90^{\circ}$ walker gets completely localized to the initial position. Therefore, from the above investigation, one can learn that for a fixed number of time steps, {\em the maximum spreading of the walker over the line can be obtained for $\theta$ values close to $0^{\circ}$ while keeping $\phi_1=\phi_2=0^{\circ}$. In contrast, when $\theta$ is close to $90^{\circ}$ and $\phi_1=\phi_2=0^{\circ}$, one can confine the walker closer to the initial position}.

\subsection{Impact of phase parameters on QW}

This section is devoted to study the impact of the phase parameters ($\phi_1$ and $\phi_2$) on the total probability distribution
for each $\theta$ value. Using Eq. (\ref{unbiased_inital_state}), we can find the total probability distribution corresponding to the initial state as follows
\begin{equation}\label{Probabilit_at_0}
P_0=|\alpha|^2+|\beta|^2    
\end{equation}
where $|\alpha|^2=1/2$ and $|\beta|^2=1/2$. By applying the general coin operator and shift operator given in (\ref{coin_operator}) and (\ref{shift_operator}) on the initial state in (\ref{unbiased_inital_state}), we can write the state at $t=1$ as follows
\begin{equation}
\begin{split}
|\bm{\psi}_1\rangle &= \alpha_1(1)|{\bf H}\rangle|\bm{1}\rangle+\beta_{\bm{-1}}(1)|{\bf T}\rangle  |{\bf -1}\rangle    
\end{split}    
\end{equation}
where
\begin{equation}\nonumber
\begin{split} 
\alpha_{\bm{1}}(1) &= \alpha \cos{\theta}+\beta e^{i\phi_1}\sin{\theta}\\
\beta_{\bm{-1}}(1) &= \alpha e^{i\phi_2}\sin{\theta}-\beta e^{i(\phi_1+\phi_2)}\cos{\theta}
\end{split}
\end{equation}
Hence, the total probability distribution at time $t=1$ is given by 
\begin{equation}\nonumber
\begin{split}
    P_1 &=(|\alpha|^2+|\beta|^2)\cos^2 \theta+(|\alpha|^2+|\beta|^2)\sin^2 \theta+\\ &\alpha\beta^{*} e^{-i\phi_1}\sin \theta \cos \theta+\alpha^*\beta e^{i\phi_1}\sin \theta \cos \theta \\
    &-\alpha\beta^* e^{-i\phi_1} \sin \theta \cos \theta -\alpha^*\beta e^{i\phi_1}\sin \theta \cos \theta
\end{split}
\end{equation}
where $\alpha^{*}$ and $\beta^{*}$ are complex conjugates. As $\alpha$ and $\beta$ are constant, probability distribution at time $t=1$ does not have any impact from the $\phi_2$. Now, 
the state of the quantum walker at an arbitrary time $k$ can be written from Eq. (\ref{Eq9}) as \begin{figure*}[!ht]
\centerline{\includegraphics[width=6in, height=3.4in]{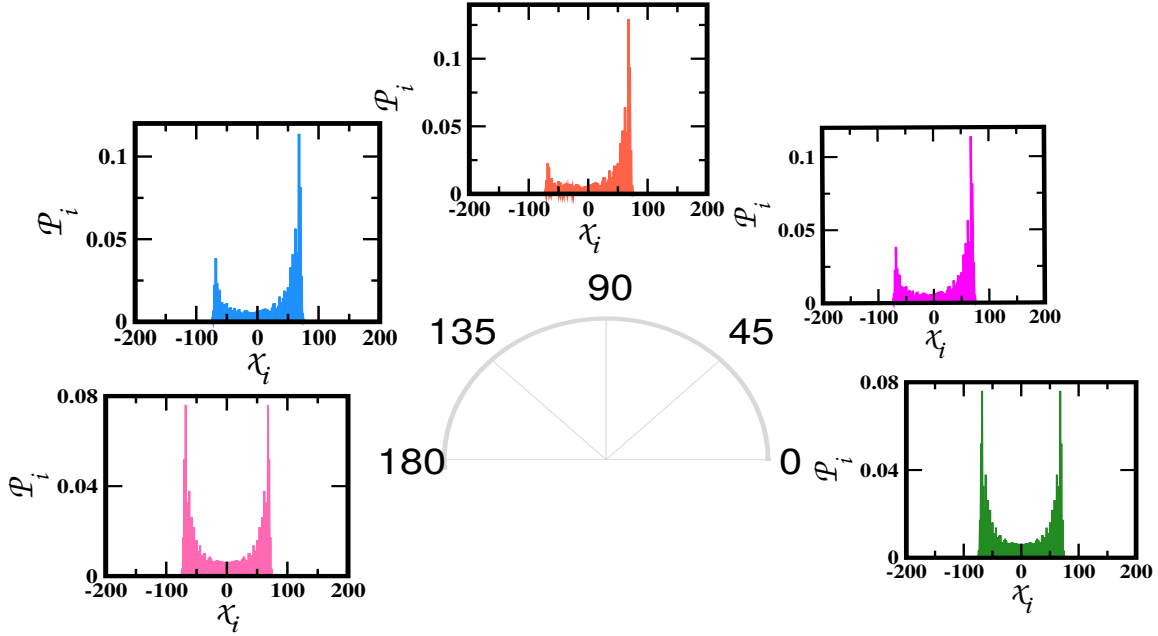}}
\caption {Impact of $\phi_1$ on the probability distribution of the QW when $\theta=45^{\circ}$. As $\phi_1$ increases from $0^{\circ}$ to $90^{\circ}$ the difference between the two peaks increases and the maximum difference appears when $\phi_1=90^{\circ}$ (distribution becomes asymmetric). The initial distribution is restored as $\phi_1$ increases from $90^{\circ}$ to $180^{\circ}$. }
\label{impact_of_phi1}
\end{figure*}

\begin{equation}\nonumber
|\bm{\psi}_k\rangle = \sum_{x \in \mathbb{Z}} (\alpha_x (k) |{\bf H}\rangle+\beta_x (k) |{\bf T}\rangle)\otimes |\bm{x}\rangle    
\end{equation}
Then, 
the corresponding probability distribution at time $k$ takes the form of (Eq. (\ref{total_probability}))
\begin{equation}\nonumber
P_k=\sum_{x \in \mathbb{Z}}{|\alpha_x(k)|^2+|\beta_x(k)|^2}    
\end{equation}
From the inductive hypothesis $P_k$ does not contain $\phi_2$.
Now, by applying the coin ($\mathcal{C}$) and shift ($\mathcal{S}$) operators given in (\ref{coin_operator}) and (\ref{shift_operator}) on $|\bm{\psi}_k\rangle$, the state at time $k+1$ can be written as 
\begin{equation}\label{total_wave_t+1}
|\bm{\psi}_{k+1}\rangle = \sum_{\bm{x} \in \mathbb{Z}} (\alpha_{\bm{x}} (k+1) |{\bf H}\rangle+\beta_{\bm{x}} (k+1) |{\bf T}\rangle)\otimes |\bm{x}\rangle    
\end{equation}
where we get the following relationship (proof in SI section 5)
\begin{equation}\label{recurrence_relation}
\begin{split}
\alpha_{\bm{x}} (k+1) &=\alpha_{\bm{x-1}} (k) \cos \theta+\beta_{\bm{x-1}} (k)e^{i\phi_1} \sin \theta\\
\beta_{\bm{x}} (k+1) & = \alpha_{\bm{x+1}} (k) e^{i\phi_2} \sin \theta-\beta_{\bm{x+1}} (k)e^{i\phi_1+\phi_2} \cos \theta \end{split}
\end{equation} 
From Eq. (\ref{total_wave_t+1}) the probability distribution at $k+1$ can be calculated as 
\begin{equation}\nonumber
\begin{split}
P_{k+1} &=\sum_{\bm{x}}{|\alpha_{\bm{x}}(k+1)|^2+|\beta_{\bm{x}}(k+1)|^2}\\
&= \sum_{\bm{x}}{\biggl(|\alpha_{\bm{x-1}}(k)|^2+|\beta_{\bm{x+1}}(k)|^2\biggr)\cos^2 \theta}\\ &
+\sum_{\bm{x}}{\biggl(|\alpha_{\bm{x+1}}(k)|^2+|\beta_{\bm{x-1}}(k)|^2\biggr)\sin^2 \theta}\\
&+\sum_{\bm{x}}\biggl(\alpha_{\bm{x-1}}(k) \beta_{\bm{x-1}}^{*}(k)e^{-i\phi_1}\\
&+\alpha_{\bm{x-1}}^{*}(k) \beta_{\bm{x-1}}(k)e^{i\phi_1}- \alpha_{\bm{x+1}}(k) \beta_{\bm{x+1}}^{*}(k)e^{-i \phi_1}\\ &-\alpha_{\bm{x+1}}^{*}(k) \beta_{\bm{x+1}}(k)e^{i \phi_1} \biggr)\sin \theta \cos \theta
\end{split}
\end{equation}
According to the expression given in (\ref{Probabilit_at_0}), the probability distribution at $t=0$ does not contain any terms of $\phi_2$. Hence, probability distribution at $t=1$ does not contain any terms of $\phi_2$. 
Thus, we can say that if the probability distribution $P_k$ at time $k$ does not contain the parameter of $\phi_2$ the distribution $P_{k+1}$ at time $k+1$ does not contain $\phi_2$ which we can observe from the induction step. 
Finally, from the principle of mathematical induction, we conclude that the total probability distribution ($P_t$) at any arbitrary time $t$ is free from $\phi_2$. The numerical verification for the inert impact of $\phi_2$ is illustrated in SI. {\em However, it is obvious that  $\theta$ and $\phi_1$ have an impact on the total probability distribution.}

Hence, we continue our study by setting $\phi_2=0^{\circ}$. Now,  we examine the impact of $\phi_1$ on the distribution in detail. Let us choose $\theta=45^{\circ}$ and vary $\phi_1$ from $0^{\circ}$ to $180^{\circ}$ (Fig. \ref{impact_of_phi1}). We wish to emphasize that the combination of $\theta=45^{\circ}$, $\phi_1=0^{\circ}$ and $\phi_2=0^{\circ}$ gives famous Hadmard coin. In Hadamard's walk, we can see two sharp peaks of equal heights (distribution is symmetric). In between the two sharp peaks, we have comparatively smaller probabilities of finding the walker. As $\phi_1$ increases from $0^{\circ}$ to $90^{\circ}$ the difference between the two sharp peaks tends to increase and reaches its maximum when $\phi_1=90^{\circ}$ (distribution becomes asymmetric). Further, as we increase $\phi_1$, the difference between the two sharp peaks tends to decrease, and when $\phi_1=180^{\circ}$, the initial distribution is restored. We observe that for each $\theta$ value, by varying $\phi_1$ we get similar behavior in the probability distribution as in Fig. \ref{impact_of_phi1}. We wish to emphasize that our analysis unveils -- for each $\phi_1\in (0,\pi)$, an asymmetry in the probability distribution even when the initial coin state is unbiased. From the above investigations, we learn that {\em rotation and one of the phase parameters play a crucial role in the final probability distribution of the walker, and by tuning them, we can regulate the dynamical behavior of the walker.}

\subsection{Algorithm for one dimensional QW}
Now, we present an algorithm for the QW on a line (Algorithm \ref{qw_algo}). We use the recurrence relation 
\begin{algorithm}[h]
\caption{1D\_Quantum\_Walk($N$, $\theta$, $\phi_1$, $\phi_2$, $\alpha$, $\beta$)} \label{qw_algo}
\begin{algorithmic}
\STATE  $n \leftarrow 2N+3$ 
\STATE $mid \leftarrow N+2$
\STATE [Coin operator]
\STATE $C_{1,1} \leftarrow \cos \theta$            
\STATE $C_{1,2} \leftarrow \sin \theta e^{i\phi_1}$
\STATE $C_{2,1} \leftarrow \sin \theta e^{i\phi_2}$            
\STATE $C_{2,2} \leftarrow -\cos \theta e^{i(\phi_1+\phi_2)}$
\STATE [Amplitude Matrix]
\STATE ${\bf A} \leftarrow \bm{0}$
\STATE ${\bf B} \leftarrow \bm{0}$
\STATE [Initial Conditions of Amplitude Matrix]
\STATE $B_{1,mid} \leftarrow \alpha$
\STATE $B_{2,mid} \leftarrow \beta$
\STATE [Amplitude Calculation]
\FOR{$k \gets 1$ to $N$}
\FOR{$m \gets 2$ to $2N+2$}
\STATE  $A_{1,m} \leftarrow C_{1,1}\times B_{1,m-1}+C_{1,2}\times B_{2,m-1}$
\STATE  $A_{2,m} \leftarrow C_{2,1} \times B_{1,m+1}+C_{2,2} \times B_{2,m+1}$
\ENDFOR
\STATE ${\bf B} \leftarrow {\bf A}$
\ENDFOR
\STATE [Probability Density Function]
\STATE $P \leftarrow \bm{0}$
\FOR{$m \gets 1$ to $n$}        
\STATE $P_m \leftarrow B_{1,m}\times B^{*}_{1,m}+ B_{2,m}\times B^{*}_{2,m}$
\STATE $x_m \leftarrow -\frac{n+1}{2}+ m$
\ENDFOR
\RETURN $P$
\end{algorithmic}
\end{algorithm}
(Eq. (\ref{recurrence_relation})) of the coefficient for the probability calculation to develop the algorithm. 
The variables $N$, $\alpha$, $\beta$, $\theta$, $\phi_1$, and $\phi_2$ are the input parameters of the algorithm. First, a positive integer value is assigned to the variable $N=t$. The arrays' size is defined by the variable $N$, and the position corresponding to the origin ($|\bm{0}\rangle$) is marked by the variable $mid$. With the help of $N$, the number of position states ($n$) and the origin is calculated. For unbiased initial coin state, we assign $\alpha=\frac{1}{\sqrt{2}}$ and $\beta=-\frac{i}{\sqrt{2}}$.  
Next, we initialize the general coin with the rotation and shift parameters, i.e., with specific values of $\theta$, $\phi_1$, and $\phi_2$. To measure the probability distribution, we store the coefficient value of both the coin states $|{\bf H}\rangle$ and $|{\bf T}\rangle$ at each position in the higher dimensional array ${\bf B}\in \mathbb{C}^{2 \times n}$ which is initialized to zero. Then the updating of coefficient value (Eq. (\ref{recurrence_relation})) in ${\bf B}$ is written in the form of a nested loop where we apply the coin to each element of the array. Finally, the probability distribution of the QW is calculated by employing the Eq. (\ref{total_probability}) where $B^{*}_{ij}$ is the complex conjugate of $B_{ij}$. The data and codes used in this paper are available at GitHub repository \cite{qw_codes_data}.

\section{Conclusion}
This article revisits the framework of discrete-time QW on a line and presents it comprehensively from an engineering perspective. We keep all the steps as simple as possible. We make the abstract notations into more concrete matrix notations. We explore the probability distribution of the discrete-time quantum walks with the general coin. We show that by fixing the initial coin state to an unbiased state and by regulating the parameters of the general coin operator, one can tune the probability distribution of the quantum walker. The general coin operator of the QWs on a line consists of three parameters (one rotation and two-phase parameters). By changing the parameters methodically, we get different coin operators. {\em Our analysis reveals that the rotation and a single phase parameter of the general coin play an essential role in controlling the walker's probability distribution. Our investigation uncovers that, for an unbiased initial coin state, we can make the probability distribution asymmetric by tuning a single phase parameter of the general coin operator.} We show that our numerical results are in good agreement with that of the analytical derivations. Among the coin operators that we have studied, Hadamard, Grover, and Fourier coins are special cases of the general coin operator. Additionally, we sketch the basic intuition of the quantum entanglement in the context of Qws. Finally, we provide an algorithm for the one-dimensional QWs.

As future extensions of this study, one can focus on (1) the dynamics of the QWs driven by the general coin operator on the graph, (2) the spectral properties of the unitary operator that comprise the general coin operator, (3) how the entanglement between coin and position states varies when the coin parameters are changed and (4) how the parameters of general coin operator can be utilized to develop tools to cater the future Engineering developments.

\ifCLASSOPTIONcompsoc

  \section*{Acknowledgment}
\fi
Mahesh N. Jayakody acknowledges the presidential scholarship of Bar-Ilan university for PhD scholars and the research funding received from Dr. Eliahu Cohen (Faculty of Engineering, Bar-Ilan University) Moreover, he is thankful to Dr. Eliahu Cohen for having useful discussions and to Prof. Asiri Nanayakkara (National Institute of Fundamental Studies, Sri Lanka) for sharing  QW code. Priodyuti Pradhan is indebted to Prof. Baruch Barzel for providing the postdoctoral research grant and acknowledges Bar-Ilan University for the Kolman-Soref postdoctoral fellowship.

\ifCLASSOPTIONcaptionsoff
  \newpage
\fi

%




\begin{thebibliography}{1}

\bibitem{Mohseni1} Mohseni, M., Read, P., Neven, H., Boixo, S., Denchev, V., Babbush, R., ... \& Martinis, J. (2017). Commercialize quantum technologies in five years. Nature News, 543(7644), 171.

\bibitem{Dowling1}Dowling, J. P., \& Milburn, G. J. (2003). Quantum technology: the second quantum revolution. Philosophical Transactions of the Royal Society of London. Series A: Mathematical, Physical and Engineering Sciences, 361(1809), 1655-1674.

\bibitem{Bresson} Bresson, A., Bidel, Y., Bouyer, P., Leone, B., Murphy, E., \& Silvestrin, P. (2006). Quantum mechanics for space applications. Applied Physics B, 84(4), 545-550.

\bibitem{Acin} Acín, Acín, A., Bloch, I., Buhrman, H., Calarco, T., Eichler, C., Eisert, J., ... \& Wilhelm, F. K. (2018). The quantum technologies roadmap: a European community view. New Journal of Physics, 20(8), 080201.

\bibitem{jayakody_2019}Jayakody, M. N., \& Nanayakkara, A. (2019). Full state revivals in higher dimensional quantum walks. Physica Scripta, 94(4), 045101.

\bibitem{Chandrashekar_qw_2020} Alderete, C. H., Singh, S., Nguyen, N. H., Zhu, D., Balu, R., Monroe, C., ... \& Linke, N. M. (2020). Quantum walks and Dirac cellular automata on a programmable trapped-ion quantum computer. Nature communications, 11(1), 1-7.

\bibitem{Wang} Wang, J., and Manouchehri, K. (2013). Physical implementation of quantum walks. Springer Berlin.

\bibitem{Quantum_algorithms_2016} Montanaro, A. (2016). Quantum algorithms: an overview. npj Quantum Information, 2(1), 1-8.

\bibitem{random_walk_quantum_walk_2020} Xia, F., Liu, J., Nie, H., Fu, Y., Wan, L., \& Kong, X. (2019). Random walks: A review of algorithms and applications. IEEE Transactions on Emerging Topics in Computational Intelligence, 4(2), 95-107.

\bibitem{Feynman} Feynman, R. P. (1986). Quantum mechanical computers. Found. Phys., 16(6), 507-532.

\bibitem{Aharonov} Aharonov, Y., Davidovich, L., and Zagury, N. (1993). Quantum random walks. Physical Review A, 48(2), 1687.

\bibitem{Kempe} Kempe, J. (2003). Quantum random walks: an introductory overview. Contemporary Physics, 44(4), 307-327.

\bibitem{Childs_2013} Childs, A. M., Gosset, D., and Webb, Z. (2013). Universal computation by multiparticle quantum walk. Science, 339(6121), 791-794.

\bibitem{qw_cryptography_2020} Abd el-Latif, A. A., Abd-el-Atty, B., Amin, M., \& Iliyasu, A. M. (2020). Quantum-inspired cascaded discrete-time quantum walks with induced chaotic dynamics and cryptographic applications. Scientific reports, 10(1), 1-16.

\bibitem{qw_complex_networks_2019} Biamonte, J., Faccin, M., \& De Domenico, M. (2019). Complex networks from classical to quantum. Communications Physics, 2(1), 1-10.

\bibitem{qw_IoT_2020} Abd El-Latif, A. A., Abd-El-Atty, B., Venegas-Andraca, S. E., Elwahsh, H., Piran, M. J., Bashir, A. K., ... \& Mazurczyk, W. (2020). Providing end-to-end security using quantum walks in IoT networks. IEEE Access, 8, 92687-92696.

\bibitem{qw_5G_communication_2020} Abd El-Latif, A. A., Abd-El-Atty, B., Mazurczyk, W., Fung, C., \& Venegas-Andraca, S. E. (2020). Secure data encryption based on quantum walks for 5G Internet of Things scenario. IEEE Transactions on Network and Service Management, 17(1), 118-131.


\bibitem{Childs_2007} Childs, A. M., Schulman, L. J., and Vazirani, U. V. (2007). Quantum algorithms for hidden nonlinear structures. In 48th Annual IEEE Symposium on Foundations of Computer Science (FOCS'07) (pp. 395-404). IEEE.

\bibitem{Ambainis} Ambainis, A. (2007). Quantum walk algorithm for element distinctness. SIAM Journal on Computing, 37(1), 210-239.

\bibitem{Magniez} Magniez, F., Santha, M., and Szegedy, M. (2007). Quantum algorithms for the triangle problem. SIAM Journal on Computing, 37(2), 413-424.

\bibitem{Hoyer} Hoyer, S., Sarovar, M., and Whaley, K. B. (2010). Limits of quantum speedup in photosynthetic light harvesting. New Journal of Physics, 12(6), 065041. 

\bibitem{Xue} Xue, P., Qin, H., and Tang, B. (2014). Trapping photons on the line: controllable dynamics of a quantum walk. Scientific reports, 4, 4825. 


\bibitem{Kitagawa_2012} Kitagawa, T., Broome, M. A., Fedrizzi, A., Rudner, M. S., Berg, E., Kassal, I., ... and White, A. G. (2012). Observation of topologically protected bound states in photonic quantum walks. Nature communications, 3, 882.

\bibitem{Douglas} Douglas, B. L., and Wang, J. B. (2008). A classical approach to the graph isomorphism problem using quantum walks. Journal of Physics A: Mathematical and Theoretical, 41(7), 075303. 

\bibitem{Strauch_2006} Strauch, F.W, (2006) Connecting the discrete and continuous-time quantum walks. Physical Review A, 74, 030301.

\bibitem{qw_graphs_2002} Aharonov, D., Ambainis, A., Kempe, J., \& Vazirani, U. (2001, July). Quantum walks on graphs. In Proceedings of the thirty-third annual ACM symposium on Theory of computing (pp. 50-59).

\bibitem{Venegas-Andraca} Venegas-Andraca, S. E. (2012). Quantum walks: a comprehensive review. Quantum Information Processing, 11(5), 1015-1106.

\bibitem{Weiss_1983} Weiss, G. H. (1983). Random walks and their applications: Widely used as mathematical models, random walks play an important role in several areas of physics, chemistry, and biology. American Scientist, 71(1), 65-71.

\bibitem{Harris_1995} Harris, E. G. (1995). Introduction to Quantum Mechanics by David J. Griffiths. AMERICAN JOURNAL OF PHYSICS, 63, 767-767.

\bibitem{Thaller} Thaller, B., 2005. {\em Advanced visual quantum mechanics}. Springer Science \& Business Media.

\bibitem{Bowers} Bowers, P.L., 2020. {\em Lectures on Quantum Mechanics: A Primer for Mathematicians}. Cambridge University Press.

\bibitem{general_coin_operator_2003} BTregenna, B., Flanagan, W., Maile, R., \& Kendon, V. (2003). Controlling discrete quantum walks: coins and initial states. New Journal of Physics, 5(1), 83.

\bibitem{Gratsea}
Gratsea, A., Metz, F., \& Busch, T. (2020). Universal and optimal coin sequences for high entanglement generation in 1D discrete time quantum walks. Journal of Physics A: Mathematical and Theoretical, 53(44), 445306.

\bibitem{Hunter} Hunter, J.K. and Nachtergaele, B., (2001). {\em Applied analysis}. World Scientific Publishing Company.

\bibitem{entanglement} Audretsch, J. (2008). {\em Entangled systems: new directions in quantum physics}. John Wiley \& Sons.

\bibitem{qw_codes_data} Our codes and data are available at the following link \url{https://github.com/priodyuti/qw_codes_data}.

































\end{thebibliography}
\end{document}


\maketitle

%
%
\newcommand{\beginsupplement}{%
\setcounter{equation}{0}
        \renewcommand{\theequation}{S\arabic{equation}}
        \setcounter{table}{0}
        \renewcommand{\thetable}{S\arabic{table}}%
        \setcounter{figure}{0}
        \renewcommand{\thefigure}{S\arabic{figure}}%
     }

\beginsupplement

\section{State of the quantum walker}
Here, we provide detail steps of the walker state calculation in Eqs. (11) and (12), where we consider $\bm{\psi}_0=|{\bf H}\rangle \otimes |\bm{x}\rangle$ and use Eqs. (6) and (7) and get 
\begin{equation}\label{Eq5}
\begin{split}
|\bm{\psi}_{1}\rangle &=\mathcal{U}|\bm{\psi}_{0}\rangle\\
&= \mathcal{S(C\otimes I)}(|{\bf H}\rangle \otimes  |\bm{x}\rangle)\\ 
&= \mathcal{S} (\mathcal{C}|{\bf H}\rangle \otimes  |\bm{x}\rangle)\\
&=\mathcal{S} \biggl(\cos \theta|{\bf H}\rangle \langle {\bf H}|{\bf H}\rangle+e^{i\phi_1}\sin \theta|{\bf H}\rangle \langle {\bf T}|{\bf H}\rangle+
     e^{i\phi_2}\sin \theta|{\bf T}\rangle \langle {\bf H}|{\bf H}\rangle-e^{i(\phi_1+\phi_2)}\cos \theta|{\bf T}\rangle \langle {\bf T}|{\bf H}\rangle \otimes  |\bm{x}\rangle \biggr)\\
&=\mathcal{S} \biggl (\cos \theta|{\bf H}\rangle  +e^{i\phi_2}\sin \theta|{\bf T}\rangle)  \otimes  |\bm{x}\rangle \biggr)\\     
&=\mathcal{S} \biggl(\cos \theta|{\bf H}\rangle \otimes  |\bm{x}\rangle  +e^{i\phi_2}\sin \theta|{\bf T}\rangle  \otimes |\bm{x}\rangle \biggr)\\
&=\biggl(\sum_{\bm{x}} |{\bf H}\rangle \langle {\bf H}|\otimes |\bm{x+1}\rangle\langle \bm{x}|+|{\bf T}\rangle \langle {\bf T}|\otimes |\bm{x-1}\rangle\langle \bm{x}|\biggr)  
 \biggl( \cos \theta|{\bf H}\rangle \otimes  |\bm{x}\rangle  +e^{i\phi_2}\sin \theta|{\bf T}\rangle  \otimes  |\bm{x}\rangle \biggr)\\
&=\biggl(\sum_{\bm{x}} (\cos \theta|{\bf H}\rangle \langle {\bf H}|{\bf H}\rangle \otimes |\bm{x+1}\rangle \langle \bm{x}|\bm{x}\rangle +
(\cos \theta|{\bf T}\rangle \langle {\bf T}|{\bf H}\rangle \otimes |\bm{x-1}\rangle\langle \bm{x}|\bm{x}\rangle)+\\
&e^{i\phi_2}\sin \theta|{\bf H}\rangle \langle {\bf H}|{\bf T}\rangle  \otimes |\bm{x+1}\rangle \langle \bm{x}|\bm{x}\rangle + e^{i\phi_2}\sin \theta|{\bf T}\rangle \langle {\bf T}|{\bf T}\rangle  \otimes |\bm{x-1}\rangle \langle \bm{x}|\bm{x}\rangle\biggr)\\
&=\cos \theta|{\bf H}\rangle  \otimes |\bm{x+1}\rangle + e^{i\phi_2}\sin \theta|{\bf T}\rangle  \otimes |\bm{x-1}\rangle
\end{split}
\end{equation}
\begin{figure}[!ht]
\centerline{\includegraphics[width=6.4in, height=1.9in]{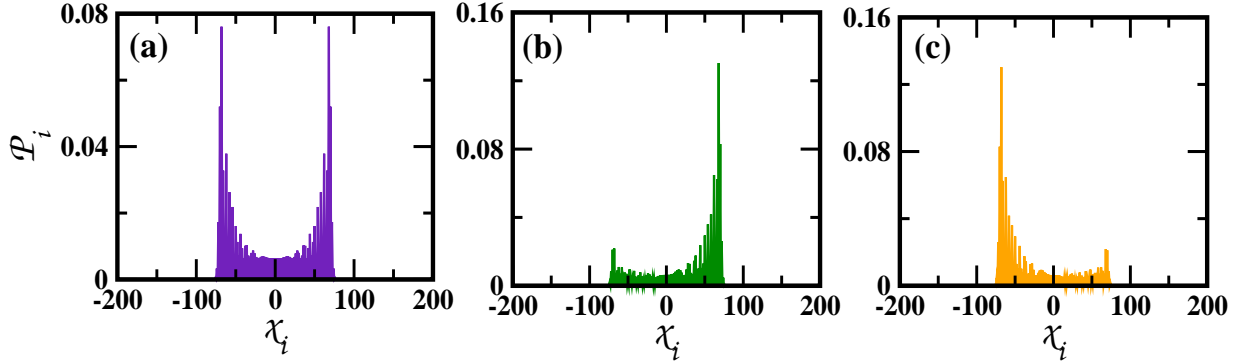}}
\caption {Compare the probability distribution of the quantum walker where we use unbiased vs. biased initial coin state ($\alpha|{\bf H}\rangle+\beta|{\bf T}\rangle$) in Eq. (18) with Hadamard coin ($\theta=45$, $\phi_1=0$, $\phi_2=0$). For $t=100$ (a) $\alpha=\frac{1}{\sqrt{2}}$ and $\beta=\frac{i}{\sqrt{2}}$ (b) $\alpha=1$ and $\beta=0$ and (c) $\alpha=0$ and $\beta=1$.}
\label{Illustrative_figures}
\end{figure}
\begin{equation}\label{Eq5}
\begin{split}
|\bm{\psi}_{2}\rangle &=\mathcal{U}|\bm{\psi}_{1}\rangle\\
&= \mathcal{S(C\otimes I)}\biggl\{cos(\theta)|{\bf H}\rangle  \otimes |\bm{x+1}\rangle + e^{i\phi_2}sin(\theta)|{\bf T}\rangle  \otimes |\bm{x-1}\rangle\biggr\}\\ 
&= \mathcal{S}\biggl\{cos(\theta)\mathcal{C}|{\bf H}\rangle  \otimes |\bm{x+1}\rangle + e^{i\phi_2}sin(\theta)\mathcal{C}|{\bf T}\rangle  \otimes |\bm{x-1}\rangle\biggr\}\\ 
&=\mathcal{S}\biggl \{\cos^2 \theta{\bf H}\rangle \langle {\bf H}|{\bf H}\rangle+e^{i\phi_1}\cos \theta \sin \theta|{\bf H}\rangle \langle {\bf T}|{\bf H}\rangle+
     e^{i\phi_2}\cos \theta \sin \theta|{\bf T}\rangle \langle {\bf H}|{\bf H}\rangle-e^{i(\phi_1+\phi_2)}\cos^2 \theta|{\bf T}\rangle \langle {\bf T}|{\bf H}\rangle \otimes  |\bm{x+1}\rangle \biggr\} \\
     &+ \mathcal{S}\biggl\{e^{i\phi_2}\sin \theta \cos \theta{\bf H}\rangle \langle {\bf H}|{\bf T}\rangle+e^{i(\phi_1+\phi_2)}\sin \theta \sin \theta|{\bf H}\rangle \langle {\bf T}|{\bf T}\rangle \\
     &+e^{2i\phi_2}\sin^2 \theta|{\bf T}\rangle \langle {\bf H}|{\bf T}\rangle-e^{i(\phi_1+2\phi_2)}\cos \theta \sin \theta|{\bf T}\rangle \langle {\bf T}|{\bf T}\rangle \otimes  |\bm{x-1}\rangle \biggr\} \\
&=\mathcal{S}\biggl\{ (\cos^2 \theta{\bf H}\rangle+ e^{i\phi_2}\cos \theta \sin \theta|{\bf T}\ ) \otimes |\bm{x+1}\rangle\} + \mathcal{S}\{(e^{i(\phi_1+\phi_2)}\sin^2 \theta|{\bf H}\rangle-e^{i(\phi_1+2\phi_2)}\cos \theta \sin \theta|{\bf T}\rangle) \otimes  |\bm{x-1}\rangle\biggr\}\\
&=\biggl(\sum_x |{\bf H}\rangle \langle {\bf H}|\otimes |\bm{x+1}\rangle\langle \bm{x}|+|{\bf T}\rangle \langle {\bf T}|\otimes |\bm{x-1}\rangle\langle \bm{x}| \biggr)\biggl(\cos^2 \theta|{\bf H}\rangle+ e^{i\phi_2}\cos \theta \sin\theta|{\bf T}\rangle \ \biggr) \otimes |\bm{x+1}\rangle\\
&+\biggl(\sum_x |{\bf H}\rangle \langle {\bf H}|\otimes |\bm{x+1}\rangle\langle \bm{x}|+|{\bf T}\rangle \langle {\bf T}|\otimes |\bm{x-1}\rangle\langle \bm{x}|\biggr)\biggl(e^{i(\phi_1+\phi_2)}\sin^2 \theta|{\bf H}\rangle-e^{i(\phi_1+2\phi_2)}\cos \theta \sin\theta|{\bf T}\rangle \biggr) \otimes  |\bm{x-1}\rangle\\
&=\cos^2 \theta|{\bf H}\rangle  \otimes |\bm{x+2}\rangle+\biggl(e^{i(\phi_1+\phi_2)}\sin^2 \theta|{\bf H}\rangle+e^{i\phi_2}\sin \theta \cos \theta|{\bf T}\rangle\biggr) \otimes  |\bm{x}\rangle -e^{i(\phi_1+2\phi_2)}\sin \theta \cos \theta|{\bf T}\rangle\otimes |\bm{x}-2\rangle
\end{split}
\end{equation}

\section{Column representation}
Here, we give column vector representation of the abstract state notations of the quantum walk. 
\begin{equation}\nonumber
    \begin{gathered}
    |\bm{\psi}_{0}\rangle
    =|\bm{H}\rangle \otimes |{\bf 0}\rangle=\begin{pmatrix}
  1 \\
  0 \\
\end{pmatrix} \otimes \begin{pmatrix}
  0 \\
  0 \\
  1 \\
  0 \\
  0 \\
\end{pmatrix}
=\begin{pmatrix}
  0 \\
  0 \\
  1 \\
  0 \\
  0 \\
  0 \\
  0 \\
  0 \\
  0 \\
  0 \\
\end{pmatrix}
\end{gathered}
\end{equation}

\begin{equation}\nonumber
   \begin{gathered}
    \begin{split}
|\bm{\psi}_{1}\rangle
&=\underbrace{\cos \theta|{\bf H}\rangle  \otimes |\bm{1}\rangle}_{|\bm{\psi}(\bm{1},1)\rangle} + \underbrace{e^{i\phi_2}\sin \theta|{\bf T}\rangle  \otimes |\bm{-1}\rangle}_{|\bm{\psi}(\bm{-1},1)\rangle}\\
&=\cos \theta\begin{pmatrix}
  1 \\
  0 \\
\end{pmatrix} \otimes \begin{pmatrix}
  0 \\
  0 \\
  0 \\
  1 \\
  0 \\
\end{pmatrix}+e^{i\phi_2}\sin \theta\begin{pmatrix}
  0 \\
  1 \\
\end{pmatrix} \otimes \begin{pmatrix}
  0 \\
  1 \\
  0 \\
  0 \\
  0 \\
\end{pmatrix}
&=\begin{pmatrix}
  0 \\
  0 \\
  0 \\
  \cos \theta \\
  0 \\
  0 \\
  e^{i\phi_2}\sin \theta \\
  0 \\
  0 \\
  0 \\
\end{pmatrix}
\end{split}
\end{gathered}
\end{equation}
\begin{equation}\nonumber
\begin{gathered}
|\bm{\psi}_{2}\rangle 
= \cos^2 \theta|{\bf H}\rangle \otimes |\bm{2}\rangle+\biggl(e^{i(\phi_1+\phi_2)}\sin^2 \theta|{\bf H}\rangle+e^{i\phi_2}\sin \theta \cos \theta|{\bf T}\rangle\biggr) \otimes|\bm{0}\rangle-e^{i(\phi_1+2\phi_2)}\sin \theta \cos \theta|{\bf T}\rangle\otimes |\bm{-2}\rangle\\
=\cos^2 \theta\begin{pmatrix}
  1 \\
  0 \\
\end{pmatrix} \otimes \begin{pmatrix}
  0 \\
  0 \\
  0 \\
  0 \\
  1 \\
\end{pmatrix}+\biggl(e^{i(\phi_1+\phi_2)}\sin^2 \theta\begin{pmatrix}
  1 \\
  0 \\
\end{pmatrix}+e^{i\phi_2}\sin \theta \cos \theta\begin{pmatrix}
  0 \\
  1 \\
\end{pmatrix}\biggr)\otimes \begin{pmatrix}
  0 \\
  0 \\
  1 \\
  0 \\
  0 \\
\end{pmatrix}-e^{i(\phi_1+2\phi_2)}\sin \theta \cos \theta\begin{pmatrix}
  0 \\
  1 \\
\end{pmatrix} \otimes \begin{pmatrix}
  1 \\
  0 \\
  0 \\
  0 \\
  0 \\
\end{pmatrix}\\
=\begin{pmatrix}
  0 \\
  0 \\
  e^{i(\phi_1+\phi_2)}\sin^2 \theta \\
  0 \\
  \cos^2 \theta \\
  -e^{i(\phi_1+2\phi_2)}\sin \theta \cos \theta  \\
  0 \\
  e^{i\phi_2}\sin \theta \cos \theta\\
  0 \\
  0\\
\end{pmatrix}
\end{gathered}
\end{equation}

\section{Quantum entanglement}
Quantum entanglement is a unique type of correlation shared between components of a quantum system \cite{Xi_2015}.
The concept of entanglement comes into the picture when we consider a system that comprises several subsystems. In other words, entanglement can be viewed as a special type of correlation between each subsystems. The mathematical framework of entanglement depends upon the fact that whether the system of interest is closed or open. That is, whether the system of interest is isolated from its environment or not. Here, we define the quantum entanglement with respect to a closed system that includes two subsystems \cite{entanglement}.
Consider two arbitrary quantum systems $A$ and $B$, with respective Hilbert spaces $\mathcal{H}_A$ and $\mathcal{H}_B$ of $d_A$ and $d_B$ dimensions. The composite Hilbert space that comprise $\mathcal{H}_A$ and $\mathcal{H}_B$ is given by $\mathcal{H}=\mathcal{H}_A\otimes \mathcal{H}_B$ with the dimension of $d=d_Ad_B$. Let $\{|\bm{a}\rangle\}_{a=0}^{d_A-1}$ and $\{|\bm{b}\rangle\}_{b=0}^{d_B-1}$ be bases of $\mathcal{H}_A$ and $\mathcal{H}_B$ respectively. Then any state of the composite system $|\bm{\phi}\rangle_{AB}\in \mathcal{H}$ can be written as 
\begin{equation}
    |\bm{\phi}\rangle_{AB}=\sum_{a=0}^{d_A-1}\sum_{b=0}^{d_B-1}{\lambda_{a,b}(|\bm{a}\rangle \otimes |\bm{b}\rangle)}
\end{equation}
where $\lambda_{a,b}\in \mathbb{C}$ and $\sum_{a=0}^{d_A-1}\sum_{b=0}^{d_B-1}{|\lambda_{a,b}|^2=1}$. Suppose there exists a set of complex numbers $\{\mu_a\}_{a=0}^{d_A-1}$ and $\{\gamma_b\}_{b=0}^{d_B-1}$ such that for each $a$ and $b$, $\lambda_{a,b}=\mu_a\gamma_b$ and $|\bm{\phi}\rangle_A=\sum_{a=0}^{d_A-1}{\mu_a|\bm{a}\rangle}$ and $|\bm{\phi}\rangle_B=\sum_{b=0}^{d_B-1}{\gamma_b|\bm{b}\rangle}$ are respective states of $\mathcal{H}_A$ and $\mathcal{H}_B$ with $\sum_{a=0}^{d_A-1}{|\mu_a|^2}=1$ and $\sum_{b=0}^{d_B-1}{|\gamma_b|^2}=1$. Then $|\bm{\phi}\rangle_{AB}$ is a separable state. Otherwise $|\bm{\phi}\rangle_{AB}$ is an entangled state. Further, if the separable condition is satisfied then $|\bm{\phi}\rangle_{AB}$  can be decomposed as $|\bm{\phi}\rangle_{AB}=|\bm{\phi}\rangle_{A} \otimes |\bm{\phi}\rangle_{B}$. For instance, consider the Hilbert spaces of $\mathcal{H}_A$ and $\mathcal{H}_B$ with two dimensions. Let $\{|\bm{0}\rangle_A,|\bm{1}\rangle_B\}$ and $\{|\bm{0}\rangle_B,|\bm{1}\rangle_B\}$ be the bases of $\mathcal{H}_A$ and $\mathcal{H}_B$ respectively. Consider a state $|\bm{\phi}\rangle_{AB} \in \mathcal{H}$ of the form  
\begin{equation}
    |\bm{\phi}\rangle_{AB}=\frac{1}{2}(|\bm{0}\rangle_A \otimes |\bm{0}\rangle_B) + \frac{1}{2}(|\bm{0}\rangle_A \otimes |\bm{1}\rangle_B) + \frac{1}{2}(|\bm{1}\rangle_A \otimes |\bm{0}\rangle_B) + \frac{1}{2}(|\bm{1}\rangle_A \otimes |\bm{1}\rangle_B)
\end{equation}
Then, we have $\lambda_{0,0}=\lambda_{0,1}=\lambda_{1,0}=\lambda_{1,1}=\frac{1}{2}$. Choose $\mu_0=\mu_1=\gamma_0=\gamma_1=\frac{1}{\sqrt{2}}$ so that $\lambda_{0,0}=\mu_0\gamma_0\text{,}\ \lambda_{0,1}=\mu_0\gamma_1\text{,}\ \lambda_{1,0}=\mu_1\gamma_0\text{,}\ \lambda_{1,1}=\mu_1\gamma_1$. Define
\begin{equation}
\begin{split}
|\bm{\phi}\rangle_A &=\frac{1}{\sqrt{2}} (|\bm{0}\rangle_A + |\bm{1}\rangle_A )\\    
|\bm{\phi}\rangle_B &=\frac{1}{\sqrt{2}} (|\bm{0}\rangle_B + |\bm{1}\rangle_B )
\end{split}    
\end{equation}
Note that $|\bm{\phi}\rangle_A \in \mathcal{H}_A$ and $|\bm{\phi}\rangle_B \in \mathcal{H}_B$. Then we have
\begin{equation}
\begin{split}
     |\bm{\phi}\rangle_A \otimes |\bm{\phi}\rangle_B&=\left (\frac{1}{\sqrt{2}} \left (|\bm{0}\rangle_A + |\bm{1}\rangle_A \right ) \right)\otimes \left (\frac{1}{\sqrt{2}} \left (|\bm{0}\rangle_B + |\bm{1}\rangle_B \right ) \right)\\
     &= \frac{1}{2}(|\bm{0}\rangle_A \otimes |\bm{0}\rangle_B) + \frac{1}{2}(|\bm{0}\rangle_A \otimes |\bm{1}\rangle_B) + \frac{1}{2}(|\bm{1}\rangle_A \otimes |\bm{0}\rangle_B) + \frac{1}{2}(|\bm{1}\rangle_A \otimes |\bm{1}\rangle_B)\\
     &=|\bm{\phi}\rangle_{AB}
\end{split}
\end{equation}
Thus $|\bm{\phi}\rangle_{AB}$ is not an entangled state but a separable state. Now consider a state of the following form
\begin{equation}
    |\bm{\psi}\rangle_{AB}=\frac{1}{\sqrt{2}}(|\bm{0}\rangle_A \otimes |\bm{0}\rangle_B) +\frac{1}{\sqrt{2}}(|\bm{1}\rangle_A \otimes |\bm{1}\rangle_B) 
\end{equation}
It is not possible to find a set of complex numbers $\{\mu_a\}$ and $\{\gamma_b\}$ such that the state $|\bm{\psi}\rangle_{AB}$ can be decomposed into two separate states from $\mathcal{H}_A$ and $\mathcal{H}_B$. Thus, $|\bm{\psi}\rangle_{AB}$ is called an entangled states.


Now we use the above definition of the entanglement to check whether we can observe entanglement in $|\bm{\psi}_{1}\rangle$ (Eq. 13). 
\begin{equation}\label{state_after_first_step}
    \begin{split}
|\bm{\psi}_{1}\rangle&=\cos \theta|{\bf H}\rangle  \otimes |\bm{1}\rangle + e^{i\phi_2}\sin \theta|{\bf T}\rangle  \otimes |\bm{-1}\rangle\\
&=\cos \theta\begin{pmatrix}
  1 \\
  0 \\
\end{pmatrix} \otimes \begin{pmatrix}
  0 \\
  0 \\
  0 \\
  1 \\
  0 \\
\end{pmatrix}+e^{i\phi_2}\sin \theta\begin{pmatrix}
  0 \\
  1 \\
\end{pmatrix} \otimes \begin{pmatrix}
  0 \\
  1 \\
  0 \\
  0 \\
  0 \\
\end{pmatrix}=\begin{pmatrix}
  0 \\
  0 \\
  0 \\
  \cos \theta \\
  0 \\
  0 \\
  e^{i\phi_2}\sin \theta \\
  0 \\
  0 \\
  0 \\
\end{pmatrix}
\end{split}
\end{equation}

Suppose $|\bm{\psi}_{1}\rangle$ is separable and there exist $|\phi_{\bm{c}}\rangle\in \mathcal{H}_{\bm{c}}$ and $|\phi_{\bm{x}}\rangle\in \mathcal{H}_{\bm{x}}$ such that 
\begin{equation}
    |\bm{\psi}_{1}\rangle=|\phi_{\bm{c}}\rangle \otimes |\phi_{\bm{x}}\rangle
\end{equation}
Now we express $|\phi_{\bm{c}}\rangle$ and $|\phi_{\bm{x}}\rangle$ in terms of the bases of their respective Hilbert spaces. Then we have
\begin{equation}
\begin{split}
     |\phi_{\bm{c}}\rangle&=\mu_1\begin{pmatrix}
  1 \\
  0 \\
\end{pmatrix} +\mu_2\begin{pmatrix}
  0 \\
  1 \\
\end{pmatrix} \\
     |\phi_{\bm{x}}\rangle&=\gamma_1\begin{pmatrix}
  1 \\
  0 \\
  0 \\
  0 \\
  0 \\
\end{pmatrix}+\gamma_2\begin{pmatrix}
  0 \\
  1 \\
  0 \\
  0 \\
  0 \\
\end{pmatrix}+\gamma_3\begin{pmatrix}
  0 \\
  0 \\
  1 \\
  0 \\
  0 \\
\end{pmatrix}+\gamma_4\begin{pmatrix}
  0 \\
  0 \\
  0 \\
  1 \\
  0 \\
\end{pmatrix}+\gamma_5\begin{pmatrix}
  0 \\
  0 \\
  0 \\
  0 \\
  1 \\
\end{pmatrix}
\end{split}
\end{equation}
where $\mu_1$,$\mu_2\in \mathbb{C}$ such that $|\mu_1|^2+|\mu_2|^2=1$ and $\gamma_1$,$\gamma_2$,$\gamma_3$,$\gamma_4$, $\gamma_5\in \mathbb{C}$ such that $|\gamma_1|^2+|\gamma_2|^2+|\gamma_3|^2+|\gamma_4|^2+|\gamma_5|^2=1$. Since we have assumed that $|\bm{\psi}_{1}\rangle$ is separable we can write
\begin{equation}\label{product_state}
    \begin{split}
        |\bm{\psi}_{1}\rangle&=\mu_1\gamma_1\begin{pmatrix}
  1 \\
  0 \\
\end{pmatrix}\otimes \begin{pmatrix}
  1 \\
  0 \\
  0 \\
  0 \\
  0 \\
\end{pmatrix} +\mu_2\gamma_1\begin{pmatrix}
  0 \\
  1 \\
\end{pmatrix}\otimes \begin{pmatrix}
  1 \\
  0 \\
  0 \\
  0 \\
  0 \\
\end{pmatrix}+\mu_1\gamma_2\begin{pmatrix}
  1 \\
  0 \\
\end{pmatrix}\otimes \begin{pmatrix}
  0 \\
  1 \\
  0 \\
  0 \\
  0 \\
\end{pmatrix}+\mu_2\gamma_2\begin{pmatrix}
  0 \\
  1 \\
\end{pmatrix}\otimes \begin{pmatrix}
  0 \\
  1 \\
  0 \\
  0 \\
  0 \\
\end{pmatrix}+\hdots
    \end{split}
\end{equation}
By comparing terms in (\ref{state_after_first_step}) and (\ref{product_state}) we get
\begin{equation}\label{terms_from_comparison}
    \begin{split}
        \mu_2\gamma_2&=\cos\theta\\
        \mu_1\gamma_4&=e^{i\phi_1}\sin\theta\\
        \gamma_1=\gamma_3&=\gamma_5=0
    \end{split}
\end{equation}
In addition, we have another two equations as follows
\begin{equation}\label{terms_form_proability}
    \begin{split}
        |\mu_1|^2+|\mu_2|^2&=1\\
       |\gamma_2|^2+|\gamma_4|^2&=1
    \end{split}
\end{equation}
From (\ref{terms_from_comparison}) one can claim that any of the $\mu_1,\ \mu_2,\ \gamma_2,\ \gamma_4$ are non-zero. Thus we have
\begin{equation}\label{seperable_state}
    \begin{split}
        |\bm{\psi}_{1}\rangle&=\mu_1\gamma_2\begin{pmatrix}
  1 \\
  0 \\
\end{pmatrix}\otimes \begin{pmatrix}
  0 \\
  1 \\
  0 \\
  0 \\
  0 \\
\end{pmatrix} +\mu_2\gamma_2\begin{pmatrix}
  0 \\
  1 \\
\end{pmatrix}\otimes \begin{pmatrix}
  0 \\
  1 \\
  0 \\
  0 \\
  0 \\
\end{pmatrix}+\mu_1\gamma_4\begin{pmatrix}
  1 \\
  0 \\
\end{pmatrix}\otimes \begin{pmatrix}
  0 \\
  0 \\
  0 \\
  1 \\
  0 \\
\end{pmatrix}+\mu_2\gamma_4\begin{pmatrix}
  0 \\
  1 \\
\end{pmatrix}\otimes \begin{pmatrix}
  0 \\
  0 \\
  0 \\
  1 \\
  0 \\
\end{pmatrix}=\begin{pmatrix}
  0 \\
  0\\
  \mu_1\gamma_2 \\
  \mu_2\gamma_2  \\
  0 \\
  0 \\
  \mu_1\gamma_4 \\
  \mu_2\gamma_4 \\
  0 \\
  0 \\
\end{pmatrix}
\end{split}
\end{equation}
But from (\ref{state_after_first_step}) we have
\begin{equation}\label{big_matrix_of_state_after_first_step}
    \begin{split}
|\bm{\psi}_{1}\rangle&=\begin{pmatrix}
  0 \\
  0 \\
  0 \\
  \cos \theta \\
  0 \\
  0 \\
  e^{i\phi_2}\sin \theta \\
  0 \\
  0 \\
  0 \\
\end{pmatrix}
\end{split}
\end{equation}
Comparing elements in (\ref{seperable_state}) and (\ref{big_matrix_of_state_after_first_step}) we get the following relations
\begin{equation}
    \begin{split}
        \mu_1\gamma_2&=0\\
        \mu_2\gamma_4&=0
    \end{split}
\end{equation}
This will lead to a contradiction. Thus, $|\bm{\psi}_{1}\rangle$ is entangled. Similarly, we can also find that $|\psi_2\rangle$ is also entangled with coin and position state. Although we start with a separable state at time $t=0$, as time progress coin and position states become more and more entangled. Several studies have been conducted to investigate quantum entanglement in the context of QWs and more details can be found in \cite{Carneiro,Abal,Singh}.

\section{Matrix formulation}
For $N=2$, we have five position states $|\bm{-2}\rangle$, $|\bm{-1}\rangle$, $|\bm{0}\rangle$, $|\bm{1}\rangle$ and  $|\bm{2}\rangle$. The coin states are given by $|\bm{H}\rangle$ and $|\bm{T}\rangle$. By using the column matrix representation of position and coin states, we can write the matrices corresponding to coin operator $\bf C$ and shift operator $\bf \bar{S}$ as follows 
\begin{equation}\label{coin_matrix}
{\bf C}=
\left( {\begin{array}{cc}
a & b \\
c & d \\
\end{array}}\right)
 \end{equation}
where $a=\cos \theta$, $b=e^{i\phi_1} \sin \theta$, $c=e^{i\phi_2}\sin \theta$ and $d=-e^{i(\phi_1+\phi_2)}\cos \theta$ such that $\theta\in [0,2\pi)$ and $\phi_1,\phi_2\in [0,\pi)$. 
\begin{equation}\label{S_for_example}
\begin{split}
{\bf \bar{S}}&=\begin{pmatrix}
  1 & 0 \\
  0 & 0 \\
\end{pmatrix}\otimes\begin{pmatrix}
  0 & 0 & 0 & 0 & 1 \\
  1 & 0 & 0 & 0 & 0 \\
  0 & 1 & 0 & 0 & 0 \\
  0 & 0 & 1 & 0 & 0 \\
  0 & 0 & 0 & 1 & 0 \\
\end{pmatrix} +\begin{pmatrix}
  0 & 0 \\
  0 & 1 \\
\end{pmatrix}\otimes \begin{pmatrix}
  0 & 1 & 0 & 0 & 0 \\
  0 & 0 & 1 & 0 & 0 \\
  0 & 0 & 0 & 1 & 0 \\
  0 & 0 & 0 & 0 & 1 \\
  1 & 0 & 0 & 0 & 0 \\
\end{pmatrix}
=\begin{pmatrix}
     0  &   0  &   0&     0&     1&     0&     0&     0&     0&     0\\
     1  &   0  &   0&     0&     0&     0&     0&     0&     0&     0\\
     0  &   1  &   0&     0&     0&     0&     0&     0&     0&     0\\
     0  &   0  &   1&     0&     0&     0&     0&     0&     0&     0\\
     0  &   0  &   0&     1&     0&     0&     0&     0&     0&     0\\
     0  &   0  &   0&     0&     0&     0&     1&     0&     0&     0\\
     0  &   0  &   0&     0&     0&     0&     0&     1&     0&     0\\
     0  &   0  &   0&     0&     0&     0&     0&     0&     1&     0\\
     0  &   0  &   0&     0&     0&     0&     0&     0&     0&     1\\
     0  &   0  &   0&     0&     0&     1&     0&     0&     0&     0
\end{pmatrix}
\end{split}
\end{equation}
By combining the matrices $\bf{C}$ and $\bf{\bar{S}}$, one can derive the unitary operator $\bf{\bar{U}}$ for the QW as follows
\begin{equation}\label{U_for_example}
\begin{split}
{\bf \bar{U}}&={\bf \bar{S}}({\bf C \otimes I})
  =\begin{pmatrix}
     0  &   0  &   0&     0&     1&     0&     0&     0&     0&     0\\
     1  &   0  &   0&     0&     0&     0&     0&     0&     0&     0\\
     0  &   1  &   0&     0&     0&     0&     0&     0&     0&     0\\
     0  &   0  &   1&     0&     0&     0&     0&     0&     0&     0\\
     0  &   0  &   0&     1&     0&     0&     0&     0&     0&     0\\
     0  &   0  &   0&     0&     0&     0&     1&     0&     0&     0\\
     0  &   0  &   0&     0&     0&     0&     0&     1&     0&     0\\
     0  &   0  &   0&     0&     0&     0&     0&     0&     1&     0\\
     0  &   0  &   0&     0&     0&     0&     0&     0&     0&     1\\
     0  &   0  &   0&     0&     0&     1&     0&     0&     0&     0
\end{pmatrix}
\left (\begin{pmatrix}
  a & b \\
c & d \\
\end{pmatrix}\otimes\begin{pmatrix}
  1 & 0 & 0 & 0 & 0 \\
  0 & 1 & 0 & 0 & 0 \\
  0 & 0 & 1 & 0 & 0 \\
  0 & 0 & 0 & 1 & 0 \\
  0 & 0 & 0 & 0 & 1 \\
\end{pmatrix}\right)\\
&=
\begin{pmatrix}
     0  &   0  &   0&     0&     1&     0&     0&     0&     0&     0\\
     1  &   0  &   0&     0&     0&     0&     0&     0&     0&     0\\
     0  &   1  &   0&     0&     0&     0&     0&     0&     0&     0\\
     0  &   0  &   1&     0&     0&     0&     0&     0&     0&     0\\
     0  &   0  &   0&     1&     0&     0&     0&     0&     0&     0\\
     0  &   0  &   0&     0&     0&     0&     1&     0&     0&     0\\
     0  &   0  &   0&     0&     0&     0&     0&     1&     0&     0\\
     0  &   0  &   0&     0&     0&     0&     0&     0&     1&     0\\
     0  &   0  &   0&     0&     0&     0&     0&     0&     0&     1\\
     0  &   0  &   0&     0&     0&     1&     0&     0&     0&     0
\end{pmatrix}
\begin{pmatrix}
a& 0& 0& 0& 0&  b&  0&  0&  0&  0\\
0& a& 0& 0& 0&  0&  b&  0&  0&  0\\
0& 0& a& 0& 0&  0&  0&  b&  0&  0\\
0& 0& 0& a& 0&  0&  0&  0&  b&  0\\
0& 0& 0& 0& a&  0&  0&  0&  0&  b\\
c& 0& 0& 0& 0& d&  0&  0&  0&  0\\
0& c& 0& 0& 0&  0& d&  0&  0&  0\\
0& 0& c& 0& 0&  0&  0& d&  0&  0\\
0& 0& 0& c& 0&  0&  0&  0& d&  0\\
0& 0& 0& 0& c&  0&  0&  0&  0& d
\end{pmatrix}\\
&=\begin{pmatrix}
0& 0& 0& 0& a&  0&  0&  0&  0&  b\\
a& 0& 0& 0& 0&  b&  0&  0&  0&  0\\
0& a& 0& 0& 0&  0&  b&  0&  0&  0\\
0& 0& a& 0& 0&  0&  0&  b&  0&  0\\
0& 0& 0& a& 0&  0&  0&  0&  b&  0\\
0& c& 0& 0& 0&  0& d&  0&  0&  0\\
0& 0& c& 0& 0&  0&  0& d&  0&  0\\
0& 0& 0& c& 0&  0&  0&  0& d&  0\\
0& 0& 0& 0& c&  0&  0&  0&  0& d\\
c& 0& 0& 0& 0& d&  0&  0&  0&  0
\end{pmatrix}
\end{split}
 \end{equation}
Now, we apply the unitary operator $\bf {\bar{U}}$ on the initial state $|\bm{\psi}_0\rangle$ given in Eq. (13) and get
\begin{equation}\label{U1_example}
\begin{split}
|\bm{\psi}_1\rangle&={\bf\bar{U}}|\bm{\psi}_0\rangle
  =\begin{pmatrix}
 0& 0& 0& 0& a&  0&  0&  0&  0&  b\\
a& 0& 0& 0& 0&  b&  0&  0&  0&  0\\
0& a& 0& 0& 0&  0&  b&  0&  0&  0\\
0& 0& a& 0& 0&  0&  0&  b&  0&  0\\
0& 0& 0& a& 0&  0&  0&  0&  b&  0\\
0& c& 0& 0& 0&  0& d&  0&  0&  0\\
0& 0& c& 0& 0&  0&  0& d&  0&  0\\
0& 0& 0& c& 0&  0&  0&  0& d&  0\\
0& 0& 0& 0& c&  0&  0&  0&  0& d\\
c& 0& 0& 0& 0& d&  0&  0&  0&  0
\end{pmatrix}
\begin{pmatrix}
  0 \\
  0 \\
  1 \\
  0 \\
  0 \\
  0 \\
  0 \\
  0 \\
  0 \\
  0 \\
\end{pmatrix}=
\begin{pmatrix}
  0 \\
  0 \\
  0 \\
  \cos \theta \\
  0 \\
  0 \\
  e^{i\phi_2}\sin \theta \\
  0 \\
  0 \\
  0 \\
\end{pmatrix}
\end{split}
\end{equation}
Further we apply ${\bf\bar{U}}$ one more time on $|\bm{\psi}_0\rangle$, and get 
\begin{equation}\label{U2_example}
\begin{split}
|\bm{\psi}_2\rangle &= {\bf {\bf\bar{U}}}^2|\bm{\psi}_0\rangle
  =\begin{pmatrix}
 bc& 0& 0& a^2&0 &  bd&  0&  0&  ab&  0\\
0& bc& 0& 0& a^2&  0&  bd&  0&  0&  ab\\
a^2& 0& bc& 0& 0&  ab&  0&  bd&  0&  0\\
0& a^2& 0& bc& 0&  0&  ab&  0&  bd&  0\\
0& 0& a^2& 0& bc&  0&  0&  ab&  0&  bd\\
ac& 0& cd& 0& 0&  bc& 0&  d^2&  0&  0\\
0& ac& 0& cd& 0&  0&  bc& 0&  d^2&  0\\
0& 0& ac& 0& cd&  0&  0&  bc& 0&  d^2\\
cd& 0& 0& ac& 0&  d^2&  0&  0&  bc& 0\\
0& cd& 0& 0& ac& 0&  d^2&  0&  0&  bc
\end{pmatrix}
\begin{pmatrix}
  0 \\
  0 \\
  1 \\
  0 \\
  0 \\
  0 \\
  0 \\
  0 \\
  0 \\
  0 \\
\end{pmatrix}=
\begin{pmatrix}
  0 \\
  0 \\
  e^{i(\phi_1+\phi_2)}\sin^2 \theta \\
  0 \\
  \cos^2 \theta \\
  -e^{i(\phi_1+2\phi_2)}\sin \theta \cos \theta  \\
  0 \\
  e^{i\phi_2}\sin \theta \cos \theta\\
  0 \\
  0\\
\end{pmatrix}
\end{split}
\end{equation}
Hence, applying ${\bf\bar{U}}$ on $|\bm{\psi}_0\rangle$ we can obtain $|\bm{\psi}_t\rangle$ for any $t$.
However, the block matrix of the unitary operator increases in size when the position space increases. Hence, we use the recursive formula in Eq. (27) to develop an algorithm (Algorithm 1) to simulate the quantum walks in practical scenarios. 

\section{Proof of the recurrence relation}
The state of the quantum walker at an arbitrary time $k$ can be written as
\begin{equation}
|\bm{\psi}_k\rangle = \sum_{\bm{x}=-\infty}^{\infty} (\alpha_{\bm{x}} (k) |{\bf H}\rangle+\beta_{\bm{x}} (k) |{\bf T}\rangle)\otimes |\bm{x}\rangle    
\end{equation}
By applying the coin ($\mathcal{C}$) and shift ($\mathcal{S}$) operators on $|\bm{\psi}_k\rangle$ in Eqs. (6) and (7), we can write the state of the quantum walker at time $k+1$ as follows

\begin{equation}\label{recurrence_formular1}
\begin{split}
|\bm{\psi}_{k+1}\rangle &=\mathcal{S}(\mathcal{C\otimes I})|\bm{\psi}_{k}\rangle\\
& =\mathcal{S}(\mathcal{C\otimes I})\left (\sum_{x=-\infty}^{\infty} (\alpha_x (k) |{\bf H}\rangle+\beta_x (k) |{\bf T}\rangle)\otimes |\bm{x}\rangle\right)\\
& = \sum_{\bm{x}=-\infty}^{\infty}(\alpha_{\bm{x}} (k) \cos \theta+\beta_{\bm{x}} (t)e^{i\phi_1} \sin \theta|{\bf H}\rangle \otimes |\bm{x+1}\rangle \\&+\sum_{\bm{x}=-\infty}^{\infty}\left(\alpha_{\bm{x}} (k) e^{i\phi_2} \sin \theta-\beta_{\bm{x}} (k)e^{i\phi_1+\phi_2} \cos \theta \right)|{\bf T}\rangle \otimes |\bm{x-1}\rangle
\end{split}
\end{equation} 
Note that the summation over $\bm{x}$ goes from $-\infty$ to $\infty$. Thus, by changing $\bm{x+1} \rightarrow \bm{x}$ and $\bm{x-1} \rightarrow \bm{x}$ we can write 

\begin{equation}\label{recurrence_formular2}
\begin{split}
|\bm{\psi}_{k+1}\rangle & = \sum_{\bm{x}=-\infty}^{\infty}\left(\alpha_{\bm{x-1}} (k) \cos \theta+\beta_{\bm{x-1}} (k)e^{i\phi_1} \sin \theta\right)|{\bf H}\rangle \otimes |\bm{x}\rangle \\&+\sum_{\bm{x}=-\infty}^{\infty}\left(\alpha_{\bm{x+1}} (k) e^{i\phi_2} \sin \theta -\beta_{\bm{x+1}} (k)e^{i\phi_1+\phi_2} \cos \theta\right)|{\bf T}\rangle \otimes |\bm{x}\rangle
\end{split}
\end{equation} 
Alternatively, the state of the quantum walker at time $k+1$ can be written in the following form as well
\begin{equation}\label{recurrence_formular3}
\begin{split}
|\bm{\psi}_{k+1}\rangle & = \sum_{x=-\infty}^{\infty} (\alpha_x (k+1) |{\bf H}\rangle+\beta_x (k+1) |{\bf T}\rangle)\otimes |\bm{x}\rangle
\end{split}
\end{equation} 
 By comparing the corresponding coefficients in (\ref{recurrence_formular2}) and (\ref{recurrence_formular3}) we can write the following recurrence formula

\begin{equation}\label{Eq19}
\begin{split}
\alpha_{\bm{x}} (k+1) &=\alpha_{\bm{x-1}} (k) \cos \theta+\beta_{\bm{x-1}} (k)e^{i\phi_1} \sin \theta\\
\beta_{\bm{x}} (k+1) & = \alpha_{\bm{x+1}} (k) e^{i\phi_2} \sin \theta-\beta_{\bm{x+1}} (k)e^{i\phi_1+\phi_2} \cos \theta \end{split}
\end{equation} 
This complete the proof.

\section{Impact of phase parameters}
\begin{figure*}[!ht]
\centerline{\includegraphics[width=4.8in, height=3.4in]{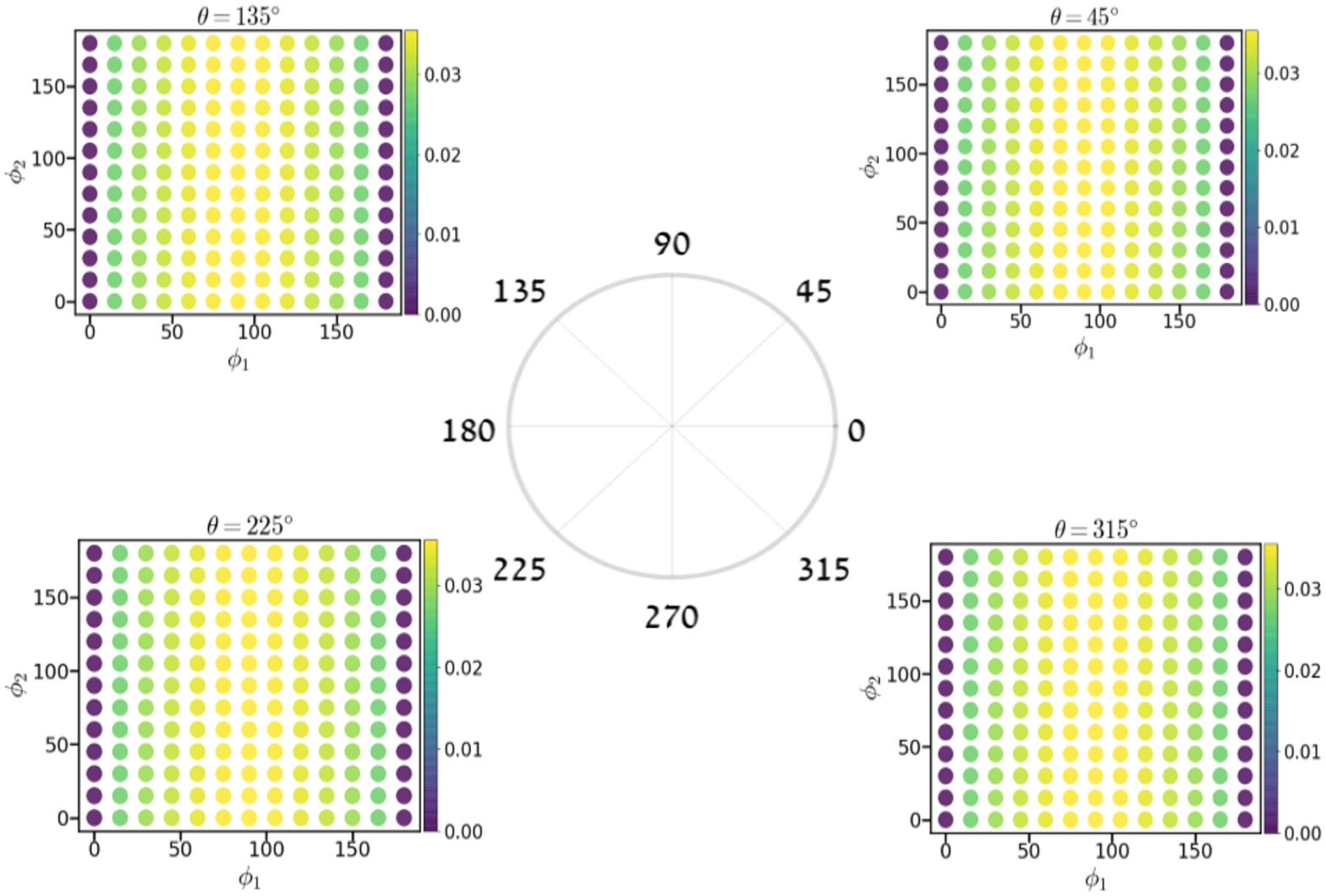}}
\caption {Phase space diagram for different $\theta \in [0,2\pi)$ and $\phi_1,\ \phi_2 \in [0,\pi)$ values when $t=100$. Shades of colours represent the magnitude of the difference between the two peaks at left and right most corners of the distribution. Colours of purple and yellow denote the minimum and maximum difference respectively. It is clearly visible that change in $\phi_2$ phase parameter has no effect on the difference but change in $\phi_1$ phase parameter has a significant impact.}
\label{phi_gradient1}
\end{figure*}

\begin{figure*}[ht]
\begin{center}
\includegraphics[width=6.4in, height=5.5in]{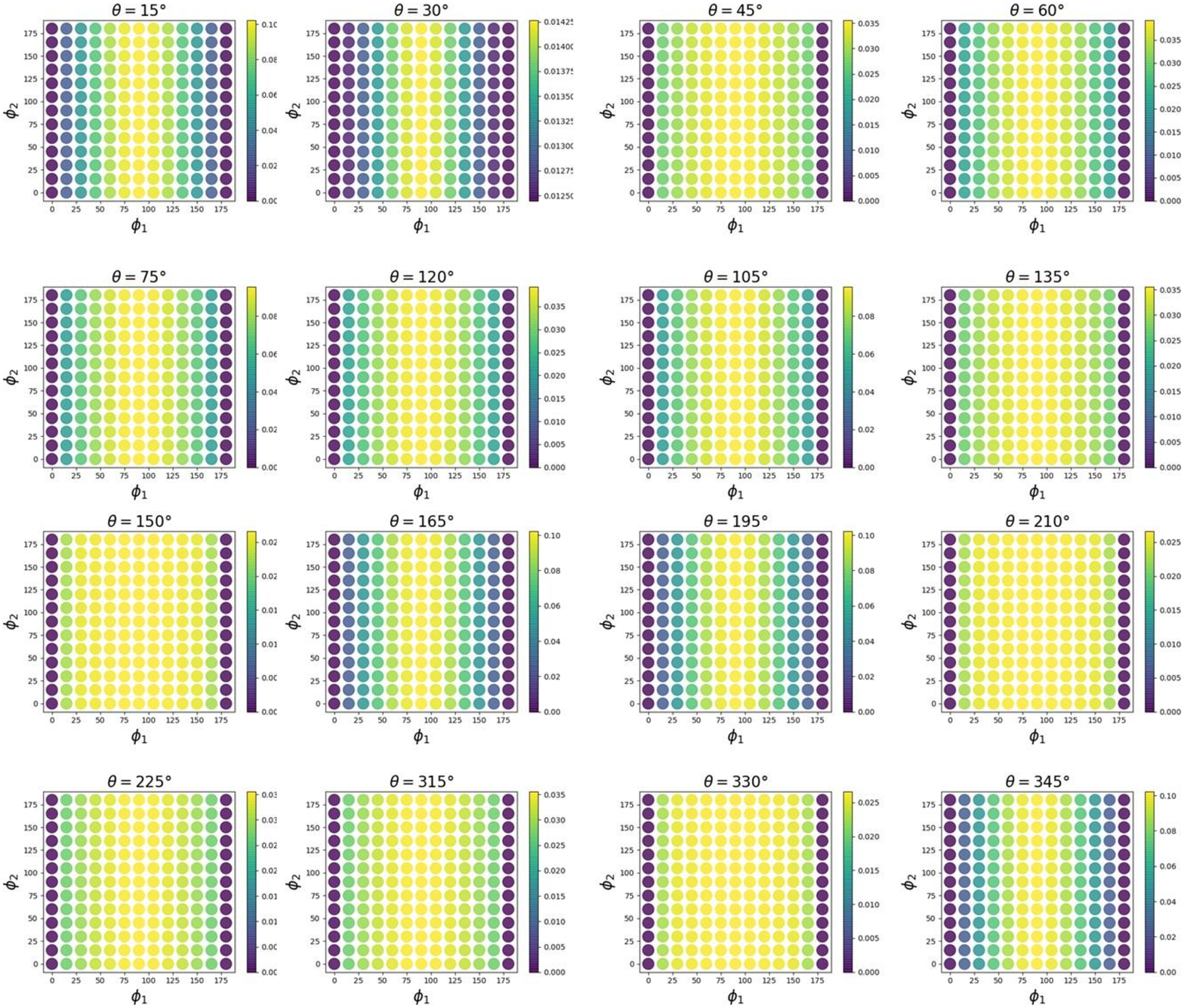}
\caption{Phase space for different $\theta$ values.} 
\label{ExampleGraphStructure}
\end{center}
\end{figure*}

If we set $\phi_1$ and $\phi_2$ to non-zero values, we can observe that the height of one peak tend to reduce and the other peak tend to increase proportionally in the probability distribution (Fig. 6). It indicates that the particle is biased towards one side of the line. Hence, we use the difference between the heights of these two peaks as a measure to quantify the impact of $\phi_1$ and $\phi_2$ on the total probability distribution of the QW. We measure the impact of $\phi_1$ and $\phi_2$ on the probability distribution as follows
\begin{equation}
    \Delta = \max_{\bm{x}} P(\bm{x},t)- \max_{\bm{y} \neq \bm{x}} P(\bm{y},t)
\end{equation}
where $P(\bm{x},t)$ is the probability at position $\bm{x}$ at time $t$ (Eq. (3)). Fig. \ref{phi_gradient1} shows a set of phase diagrams from which we can extract the information about the impact of $\phi_1$ and $\phi_2$ on the distribution. 
From Fig. \ref{phi_gradient1}, one can observe that the parameter $\phi_2$ has no impact on the probability distribution. That is, when we vary $\phi_2$ while keeping $\phi_1$ for a fixed value, there are no difference in the shape of the total probability distribution. 
However, the coin parameter $\phi_1$ has a significant impact on the shape of the probability distribution. The phase diagram for other $\theta$ values can be shown in Fig. \ref{ExampleGraphStructure}.
